\documentclass[aps,prd,reprint,superscriptaddress,nofootinbib]{revtex4-2}
\usepackage{amsmath, amsfonts}
\usepackage{graphicx}
\usepackage{bm}
\usepackage{ulem}
\usepackage{color}
\usepackage[dvipsnames]{xcolor}
\usepackage{booktabs}
\usepackage[colorlinks=true,pdfstartview=FitV,linkcolor=blue,citecolor=blue,urlcolor=blue]{hyperref}

\enlargethispage{\baselineskip}

\begin{document}

\title{Primordial Black Hole Abundance from Reionization}

\newcommand{\TDLI}{\affiliation{Tsung-Dao Lee Institute \& School of Physics and Astronomy, Shanghai Jiao Tong University, Shanghai 201210, China}}

\author{Ziwen Yin}
\email{ziwenyin@sjtu.edu.cn}
\TDLI

\author{Hanyu Cheng}
\email{chenghanyu@sjtu.edu.cn}
\TDLI

\author{Luca Visinelli}
\email{lvisinelli@unisa.it}
\affiliation{Dipartimento di Fisica ``E.R.\ Caianiello'', Universit\`a degli Studi di Salerno,\\ Via Giovanni Paolo II, 132 - 84084 Fisciano (SA), Italy}
\affiliation{Istituto Nazionale di Fisica Nucleare - Gruppo Collegato di Salerno - Sezione di Napoli,\\ Via Giovanni Paolo II, 132 - 84084 Fisciano (SA), Italy}

\begin{abstract}
We derive robust constraints on the initial abundance of evaporating primordial black holes (PBHs) using the reionization history of the Universe. We consider PBHs with initial masses $3.2\times10^{13},\mathrm{g}\lesssim M_{\rm PBH}\lesssim5\times10^{14},\mathrm{g}$, whose Hawking emission injects electromagnetic (EM) energy into the intergalactic medium (IGM) after recombination. Using \texttt{BlackHawk}, we compute the redshift-dependent emission spectra, accounting for PBH mass evolution and the full spectrum of Standard Model particles and gravitons. The photon and electron spectra are folded with channel-specific \texttt{DarkHistory} deposition efficiencies, which encode EM cascade cooling and delayed deposition into hydrogen ionization and excitation and IGM heating. The modified ionization histories determine the high-redshift Thomson optical depth, which we constrain using the posterior obtained from a Gaussian-process reconstruction of the free-electron fraction based on low-$\ell$ \textit{Planck} CMB polarization data. This nonparametric reconstruction provides an operational separation between high-redshift energy injection and late-time astrophysical reionization without imposing a fixed ionization template. Requiring consistency with this posterior, we obtain upper limits on the initial PBH abundance that are less sensitive to late-time reionization assumptions. Compared with earlier analytic CMB estimates, our bounds are generally more conservative, reflecting the restriction to low-$\ell$ polarization data and the time-dependent treatment of PBH mass evolution and Hawking emission combined with channel- and redshift-dependent deposition efficiencies. Our results demonstrate the power of reionization observables as a precision probe of PBH evaporation.
\end{abstract}

\maketitle

\section{Introduction}
\label{sec:introduction}

Black holes (BHs) are among the most striking predictions of general relativity (GR), representing regions of spacetime where gravitational collapse leads to the formation of an event horizon. Over the past decades, compelling observational evidence has established their existence across a wide range of mass scales, including measurements of stellar orbits around Sagittarius~A$^\star$ at the Galactic Center~\cite{Ghez:1998ph,Ghez:2008ms,Gillessen:2008qv}, detections of gravitational waves (GW) from binary BH mergers~\cite{LIGOScientific:2016aoc, LIGOScientific:2018mvr}, and horizon-scale imaging of supermassive BHs in nearby galaxies~\cite{EventHorizonTelescope:2019dse, EventHorizonTelescope:2022xnr}. In the standard astrophysical picture, BHs form from the gravitational collapse of massive stars at the end of their life cycles~\cite{Heger:2002by}.

In addition to these astrophysical objects, BHs may also have formed in the early Universe as primordial black holes (PBHs), generated by the collapse of large density fluctuations shortly after horizon re-entry~\cite{Zeldovich:1967lct, Carr:1974nx, Carr:1975qj}. Such fluctuations can arise in a variety of well-motivated scenarios, including enhanced curvature perturbations generated during inflation~\cite{Garcia-Bellido:1996mdl}, modifications of the primordial equation of state~\cite{Byrnes:2018clq}, first-order phase transitions~\cite{Maeso:2021xvl, Wang:2025hwc}, scalar-mediated long-range forces~\cite{Flores:2020drq}, or bubble collisions in the early Universe~\cite{Hawking:1982ga, Kodama:1982sf}. The abundance of PBHs at formation is commonly estimated using Press–Schechter or peak-theory approaches, which require the density contrast to exceed a critical threshold $\delta_c \approx 0.4-0.7$~\cite{Shibata:1999zs, Niemeyer:1999ak, Musco:2004ak, Wu:2020ilx, Gow:2020bzo}. Comprehensive reviews of PBH formation and phenomenology can be found in Refs.~\cite{Carr:2009jm, Carr:2016drx, Carr:2020gox, Carr:2020xqk, Green:2020jor}.

BHs are expected to emit Hawking radiation due to quantum effects near the event horizon~\cite{Hawking:1974rv}. As a consequence, PBHs lose mass and eventually evaporate, with a lifetime that depends sensitively on their initial mass. PBHs heavier than $M_{\rm PBH} \gtrsim 5\times 10^{14}$\,g survive until the present epoch and have been extensively studied as potential dark matter (DM) candidates, constrained by microlensing~\cite{Niikura:2017zjd}, dynamical effects~\cite{Carr:1997cn, Carr:2020erq}, and cosmic microwave background (CMB) observations~\cite{Ali-Haimoud:2016mbv, Serpico:2020ehh}, depending on their mass window. PBHs in the range $M_{\rm PBH} \sim (10^{15}\textrm{--}10^{17})$\,g are additionally constrained by searches for gamma rays produced during their evaporation~\cite{Coogan:2020tuf, Ray:2021mxu, Keith:2022sow}. PBHs with masses below $\sim 5\times 10^{14}\,\mathrm{g}$ instead evaporate before the present epoch, injecting energy into the intergalactic medium after recombination and before or during the epoch of reionization. In this regime, gamma-ray constraints rapidly weaken, while the ionization and thermal histories of the intergalactic medium (IGM) provide a sensitive probe of PBH evaporation. Astrophysical uncertainties in the timing and duration of reionization, however, can substantially weaken such bounds unless they are treated in a model-independent and data-driven manner. PBH evaporation gives rise to a rich phenomenology, including observable cosmological signatures~\cite{Auffinger:2022khh, Korwar:2023kpy, Mazde:2022sdx, Vitale:2026ric}, the production of dark matter and dark radiation through Hawking emission~\cite{Fujita:2014hha, Allahverdi:2017sks, Lennon:2017tqq, Morrison:2018xla, Hooper:2019gtx, Baldes:2020nuv, Masina:2020xhk, Gondolo:2020uqv, Cheek:2021cfe, Chattopadhyay:2022fwa}, and significant implications for axion-like particle cosmology~\cite{Bernal:2021bbv, Schiavone:2021imu}. In addition, the interplay between PBHs and DM halos surrounding them can lead to a variety of constraints on black hole and DM formation models~\cite{Boucenna:2017ghj, Bertone:2019vsk, Hertzberg:2020hsz, Carr:2020mqm, Gines:2022qzy, Chanda:2022hls, Bertone:2024wbn, Jangra:2024sif, Boccia:2024nly, Khan:2025kag}. While related effects often probe PBHs evaporating before or during Big-Bang nucleosynthesis, in this work we focus on later-evaporating PBHs whose dominant cosmological impact arises from energy injection into the IGM after recombination.

In this work, we investigate the cosmological effects of PBHs with initial masses $M_{\rm PBH}\sim(10^{13}\textrm{--}10^{14})\,\mathrm{g}$, whose Hawking evaporation injects electromagnetic (EM) energy into the IGM predominantly after recombination. Depending on their mass, these PBHs either evaporate completely during the post-recombination epoch or remain partially unevaporated until the present day. We restrict our analysis to initial abundances for which the PBH population remains cosmologically subdominant and does not generate an early matter-dominated era. Instead, their cosmological signatures arise from late-time energy injection through Hawking radiation, leading to ionization, excitation, and heating of hydrogen and helium during the post-recombination epoch. These processes modify the ionization history and leave observable imprints on precision cosmological observables. Energy injection from evaporating PBHs has been previously constrained using CMB anisotropies~\cite{Chen:2003gz, Padmanabhan:2005es, Slatyer:2009yq, Slatyer:2015jla, Poulin:2016anj, Stocker:2018avm, Poulter:2019ooo, Yang:2019bkk}, CMB spectral distortions~\cite{Tashiro:2008sf, Chluba:2013dna, Chluba:2019nxa, Lucca:2019rxf, Acharya:2019xla, Acharya:2020jbv}, and global 21-cm measurements~\cite{Clark:2018ghm, Mirocha:2020slz}. Many of these analyses rely on semi-analytical treatments of the energy injection history or assume parametric models of reionization, which can obscure the separation between exotic energy injection and astrophysical reionization processes. Recently, a robust, data-driven framework for constraining late-time energy injection using reionization observables has been developed~\cite{Cheng:2025cmb}, in which the free-electron fraction is reconstructed directly from low-$\ell$ CMB polarization data using Gaussian Process (GP) regression. The corresponding numerical implementation is publicly available at~\href{https://github.com/Cheng-Hanyu/CLASS_reio_gpr}{github.com/Cheng-Hanyu/CLASS\_reio\_gpr}. This framework constrains exotic energy injection scenarios without imposing a fixed parametric form for the high-redshift ionization history.

The goal of this work is to apply this reionization methodology to evaporating PBHs in the mass range $(10^{13}\textrm{--}10^{14})\,\mathrm{g}$. We model Hawking evaporation including the full set of relevant Standard Model degrees of freedom, track the resulting time-dependent energy injection and EM cascades in the IGM, and derive constraints on the initial PBH abundance, commonly parametrized by the mass fraction at formation $\beta$. Our analysis follows a strategy similar to that developed for ensembles of decaying relics~\cite{Yin:2025amn}, adapted here to the intrinsically time-dependent nature of PBH evaporation. We show that reionization observables provide competitive and complementary constraints in this mass window, highlighting the unique sensitivity of the ionization history to late-time energy injection.

The paper is organized as follows. In Sec.~\ref{sec:methods} we review the formation and evaporation of PBHs, describe the computation of their Hawking emission spectra, and outline the cosmological framework used to model energy injection and EM cascades in the IGM. In Sec.~\ref{sec:analysis} we present the analysis pipeline, detailing the treatment of the ionization and thermal histories, the GP reconstruction of the reionization history, and the extraction of constraints from the high-redshift optical depth. Our results are presented in Sec.~\ref{sec:results}, where we derive bounds on the initial PBH abundance as a function of mass, and are discussed in Sec.~\ref{sec:discussion} through comparisons with existing constraints and an assessment of the physical effects. We summarize our findings and outline future prospects in Sec.~\ref{sec:conclusions}. Throughout this work we use natural units $\hbar = c = k_B = 1$, and we refer to the Planck mass in terms of Newton’s constant $G_N$ as $m_{\rm Pl} = G_N^{-1/2}$.

\section{Methods}
\label{sec:methods}

PBHs are characterized by their initial abundance, typically defined as the fraction of the Universe’s energy density that collapses into black holes at the time of formation $t_f$,
\begin{equation}
    \label{eq:definebeta}
    \beta = \frac{8\pi G_N}{3 H^2(t_f)}\,\rho_{\rm PBH}(t_f)\,,
\end{equation}
where $\rho_{\rm PBH}$ is the PBH energy density and $H(t)$ is the Hubble rate at time $t$. In the absence of evaporation, the PBH energy density scales as matter and grows relative to radiation until matter-radiation equality. It is convenient to relate $\beta$ to an effective PBH fraction of dark matter at equality, $f_{\rm PBH}$, using entropy conservation during radiation domination,
\begin{equation}
    \beta = f_{\rm PBH}\,\left(\frac{g_{*s}(T_{\rm eq})}{g_{*s}(T_f)}\right)^{1/3} \frac{T_{\rm eq}}{T_f}\,.
    \label{eq:fPBH-beta-eq}
\end{equation}
Throughout this work we quote constraints in terms of $\beta$, which remains well-defined even when PBHs evaporate before matter-radiation equality.

In the standard scenario, PBHs form during radiation domination from the collapse of overdense regions~\cite{Carr:1974nx, Carr:1975qj}. Assuming a monochromatic mass function, the PBH mass at formation is related to the horizon mass by a collapse efficiency factor $\gamma \simeq 0.2$,
\begin{equation}
    \label{eq:initialmass}
    M_{\rm PBH} = \gamma \frac{4\pi}{3}\frac{\rho_{\rm crit}(t_f)}{H_f^3} = \frac{\gamma}{2G_N H_f}\,,
\end{equation}
where $H_f \equiv H(t_f)$. In our analysis, the PBH mass $M_{\rm PBH}$ and initial abundance $\beta$ are treated as independent parameters.

Prior to their complete evaporation, the physical number density of PBHs at redshift $z$, expressed in terms of the effective fraction $f_{\rm PBH}$, is
\begin{equation}
    n_{\rm PBH}(M_{\rm PBH},z) = f_{\rm PBH}\,\frac{\rho_c\,\Omega_{\rm DM}}{M_{\rm PBH}}\,(1+z)^3\,,
\end{equation}
where $\rho_c \equiv 3 H_0^2/(8\pi G_N)$ is the present-day critical density, which depends on the value of the Hubble constant $H_0$, and $\Omega_{\rm DM}$ is the present-day dark matter density parameter. For the PBH masses considered in this work, however, Hawking evaporation proceeds over cosmological timescales, and the PBH mass evolves appreciably between recombination and reionization. PBHs lose mass through Hawking evaporation, emitting all particle species with masses below the Hawking temperature~\cite{Hawking:1975vcx, Page:1976wx, Carr:1976zz}
\begin{equation}
    \label{eq:Hawking}
    T_H(t) = \frac{1}{8\pi G_N M_{\rm PBH}(t)}\,.
\end{equation}
We assume that PBHs with masses $M_{\rm PBH} \sim (10^{13}\textrm{--}10^{14})\,\mathrm{g}$, formed from near-spherical collapse during radiation domination, are well approximated as non-spinning~\cite{Harada:2017fjm,DeLuca:2019buf}. The total mass loss rate of a non-rotating PBH is given by
\begin{equation}
    \label{eq:massloss}
    \frac{{\rm d}M_{\rm PBH}}{{\rm d}t} = -\sum_i \int {\rm d}E\, E\, \frac{{\rm d}^2 N_i}{{\rm d}E\,{\rm d}t} \!\left(E; M_{\rm PBH}(t)\right)\,,
\end{equation}
where the sum runs over all particle species $i$. The emission rate for the $i$-th degrees of freedom with spin $s_i$ and internal degrees of freedom $g_i$, per unit time and energy is given as
\begin{equation}
    \frac{{\rm d}^2 N_i}{{\rm d}E\,{\rm d}t}\!\left(E; M_{\rm PBH}(t)\right) = \sum_{\ell} \frac{g_i}{2\pi} \frac{\Gamma_{s_i,\ell}(E)}{e^{E/T_H} \pm 1}\,,
\end{equation}
where $\Gamma_{s_i,\ell}(E)$ denotes the graybody factor encoding the probability for a mode of energy $E$ and angular momentum $\ell$ to escape the BH potential barrier~\cite{Teukolsky:1973ha}. For non-rotating PBHs, these graybody factors are obtained by solving the Bardeen-Press equation for each spin~\cite{Bardeen:1973xb, Page:1976wx, MacGibbon:1990zk, MacGibbon:1991tj}. While the PBH lifetime scales parametrically as $\tau\propto M_{\rm PBH}^3$, accurate modeling of the evaporation history requires numerical integration of the mass loss rate, particularly near the end of the PBH lifetime when the emission rate increases rapidly.

In this work, we compute the full time evolution of the PBH mass by numerically evolving the mass loss Eq.~\eqref{eq:massloss} using the \texttt{BlackHawk 2.3} code~\cite{Arbey:2019mbc, Arbey:2021mbl, Auffinger:2022sqj}, including all relevant SM degrees of freedom as well as gravitons. This fully time-dependent treatment is essential for accurately modeling the redshift distribution of energy deposition, particularly near the end of the PBH lifetime, when the emission rate increases rapidly and can significantly affect the ionization history of the IGM. The instantaneous PBH mass $M_{\rm PBH}(t)$ is used to evaluate the emitted particle spectra and energy injection rate at each redshift. Primary spectra for photons and electrons/positrons are obtained directly from \texttt{BlackHawk 2.3}, while secondary EM emission from particle decays and final-state radiation is computed using \texttt{HAZMA}~\cite{Coogan:2019qpu}, a code designed to model EM emission from particle injection in astrophysical environments at late times. Given the emission rate for photons and electrons, the total injected EM energy density per unit physical volume and time is
\begin{equation}
    \label{eq:Einj}
    \left(\frac{{\rm d}E}{{\rm d}V\,{\rm d}t}\right)_{\rm inj}\!\!(z) = n_{\rm PBH}\!\!\sum_{i\in\{\gamma,e^\pm\}} \!\int {\rm d}E E \frac{{\rm d}^2 N_i}{{\rm d}E\,{\rm d}t} \!\left(E; M_{\rm PBH}(t)\right).
\end{equation}
Injected photons and electrons initiate EM cascades through Compton scattering, inverse Compton scattering, pair production, and photoionization. Energy deposition is not instantaneous and is described using redshift-dependent deposition efficiencies. We take the deposited power per unit time to be positive when energy is transferred to the IGM. In redshift space, we define
\begin{equation}
	 \left.\frac{{\rm d}E}{{\rm d}V\,{\rm d}z}\right|_{{\rm dep},c} = \frac{{\rm d}t}{{\rm d}z}\left.\frac{{\rm d}E}{{\rm d}V\,{\rm d}t}\right|_{{\rm dep},c}\,, \qquad
	\frac{{\rm d}t}{{\rm d}z} = -\frac{1}{(1+z)H(z)}\,.
\end{equation}
Consequently, positive energy deposition per unit time corresponds to a negative signed derivative with respect to redshift.

\section{Analysis}
\label{sec:analysis}

\subsection{Ionization of the IGM by PBH evaporation}
\label{sec:reionization}

We analyze the impact of PBH Hawking evaporation on the thermal and ionization history of the IGM between recombination and the onset of astrophysical reionization. In this regime the ionization history is primarily governed by standard recombination physics, but it can be modified by additional sources of energy injection such as the EM products of PBH evaporation. During this epoch, the free-electron fraction evolves under the combined action of recombination processes and any additional injected energy. The ionization state of the IGM is characterized by the ionization fraction
\begin{equation}
    \label{eq:Xe_def}
    X_e \equiv \frac{n_e}{n_H}\,,
\end{equation}
where $n_e$ is the number density of free electrons and $n_H$ is the total number density of hydrogen nuclei. The kinetic temperature of the baryonic gas $T_b$ is defined in terms of the mean thermal energy per particle,
\begin{equation}
    \frac{3}{2}\,k_B\,T_b = \langle\frac{1}{2}m_bv^2\rangle\,,
\end{equation}
where $m_b$ is the particle mass and the angular brackets denote an average over its thermal velocity distribution. The evolution of $X_e$ and $T_b$ is described by the coupled differential equations~\cite{Slatyer:2016qyl, Liu:2019bbm, Escudero:2023vgv}
\begin{align}
    &\frac{{\rm d}X_e}{{\rm d}z} \label{eq:Xe}\\
    = &\frac{C_H}{(1+z)H(z)}\!\left(\!-\beta_H(T_\gamma)(1 \!-\! X_e)e^{\frac{-E_{H,2s1s}}{T_\gamma}} \!\! +\! X_e^2n_H\alpha_H(T_b)\!\right)\notag\\
    &+\frac{1}{n_HE_i}\frac{{\rm d}E}{{\rm d}V{\rm d}z}\bigg|_{\mathrm{dep,ion}}\!\!\!\!+(1-C_H)\frac{1}{n_HE_{H,1s2p}}\frac{{\rm d}E}{{\rm d}V{\rm d}z}\bigg|_{\mathrm{dep,exc}}\!\!,\notag \\
    &(1+z)\frac{{\rm d}T_b}{{\rm d}z}=2T_b+\frac{8}{3}\frac{\rho_\gamma\sigma_T}{m_e\,H(z)}\frac{X_e}{1+f_{\rm He}+X_e}(T_b-T_\gamma)\notag\\
    &+\frac{2}{3}\frac{1+z}{n_H(1+f_{\rm He}+X_e)}\frac{{\rm d}E}{{\rm d}V{\rm d}z}\bigg{|}_\mathrm{dep,heat}\,. \label{eq:Tb}
\end{align}
Here, $T_\gamma$ is the photon temperature, related to the photon energy density by $\rho_\gamma = (\pi^2/15)T_\gamma^4$, and $\sigma_T$ is the Thomson cross section. The coefficients $\alpha_H$ and $\beta_H$ denote the case-B recombination and photoionization rates, respectively. The hydrogen excitation and ionization energies are $E_{H,1s2p}=E_{H,2s1s}=10.2\,\mathrm{eV}$ and $E_i=13.6\,\mathrm{eV}$, and $f_{\rm He}$ is the helium-to-hydrogen number density ratio. The Peebles factor $C_H$ accounts for the probability that an excited hydrogen atom reaches the ground state before being ionized. At high redshift, baryons remain tightly coupled to photons through efficient Compton scattering, so that $T_b = T_\gamma$. After recombination, Compton coupling weakens and the baryon temperature evolves approximately adiabatically, with $T_\gamma \propto (1+z)$ and $T_b \propto (1+z)^2$, leading to a decreasing ratio $T_b/T_\gamma$ around $z \sim 200\textrm{--}1000$. Any additional heating source such as PBH evaporation modifies this post-recombination thermal evolution through the extra terms in Eqs.~\eqref{eq:Xe}--\eqref{eq:Tb}.

Energy injected into the IGM by PBH evaporation is redistributed among ionization, excitation, and heating channels, labeled by $c=\{\mathrm{ion},\mathrm{exc},\mathrm{heat}\}$~\cite{Chen:2003gz,
Padmanabhan:2005es,Slatyer:2009yq,Slatyer:2015jla,Poulin:2016anj}. The deposited energy rate per unit time in each channel is
\begin{equation}
    \label{eq:Edep-fc}
    \frac{{\rm d}E}{{\rm d}V\,{\rm d}t}\bigg{|}_{{\rm dep},c}\!\!(z) = f_{c}(z)\,\left(\frac{{\rm d}E}{{\rm d}V\,{\rm d}t}\right)_{\rm inj}\!\!(z)\,,
\end{equation}
where $f_c(z)$ is the channel-specific deposition efficiency. We evaluate the efficiencies for hydrogen ionization, excitation, and heating separately using the tabulated \texttt{DarkHistory} deposition functions~\cite{Liu:2019bbm}. For the broad PBH emission spectrum, the effective channel efficiency is constructed as
\begin{equation}
    \label{eq:fc}
    f_c(z)= \frac{\sum_{i\in\{\gamma,e^\pm\}}\int {\rm d}E\,E\,\frac{{\rm d}^2N_i}{{\rm d}E\,{\rm d}t}\!\left(E; M_{\rm PBH}(z)\right)f_{c,i}(E,z)}{\displaystyle\sum_{i\in\{\gamma,e^\pm\}}\int {\rm d}E\,E\,\frac{{\rm d}^2N_i}{{\rm d}E\,{\rm d}t}\!\left(E; M_{\rm PBH}(z)\right)}\,,
\end{equation}
where $f_{c,i}(E,z)$ is the tabulated channel-specific deposition efficiency for an injected particle of species $i$ and energy $E$. These efficiencies encode the non-instantaneous energy deposition associated with EM cooling cascades. The resulting channel-resolved deposited powers enter as source terms in Eqs.~\eqref{eq:Xe}--\eqref{eq:Tb}. The photon temperature evolves as
\begin{equation}
    T_\gamma = T_{\gamma0}\,(1+z)\,.
\end{equation}

Neglecting the free-electron contribution from ionized helium, the total hydrogen-nucleus number density is
\begin{equation}
    n_H = n_{\rm HI}+n_{\rm HII} = 3.1\times10^{-8}\,\eta_{10}\left(\frac{T_\gamma}{T_{\gamma0}}\right)^3\,\mathrm{cm}^{-3}\,,
\end{equation}
where $n_{\rm HI}$ and $n_{\rm HII}\simeq n_e$ are the neutral- and ionized-hydrogen number densities, respectively, $\eta_{10}=10^{10}n_b/n_\gamma\simeq6.1$ is the baryon-to-photon ratio~\cite{Fields:2019pfx}, and $T_{\gamma0}\simeq2.725\,\mathrm{K}$ is the present-day photon temperature. Computing the total optical depth requires extending the ionization history to low redshift. The helium-to-hydrogen number density ratio is
\begin{equation}
    f_{\rm He} = \frac{n_{\mathrm{He}}}{n_H} = \frac{Y_p}{4(1-Y_p)}\,,
\end{equation}
where $Y_p$ is the primordial helium mass fraction. At $z\lesssim30$, we supplement the modified residual ionization history with the standard one-step hyperbolic-tangent prescription implemented in \texttt{CAMB}~\cite{Lewis:2008wr,Lewis:2026mif},
\begin{equation}
    \label{eq:reionization}
    X_{\rm rei}(z) = \frac{1+f_{\rm He}}{2}\left[1+\tanh\left(\frac{(1+z_{\rm re})^{3/2}-(1+z)^{3/2}}{(3/2)(1+z_{\rm re})^{1/2}\Delta z}\right)\right]\,,
\end{equation}
with $z_{\rm re}=6.1$ and $\Delta z=0.5$. The full ionization history is obtained by adding the solution of the modified recombination equations, including PBH energy injection, to the phenomenological contribution from late-time astrophysical reionization. The free-electron fraction entering the optical depth calculation in Eq.~\eqref{eq:opticaldepth} is
\begin{equation}
    X_e^{\rm tot}(z)=X_e(z)+X_{\rm rei}(z)\,.
\end{equation}
At redshifts well above reionization, $X_{\rm rei}(z)\to0$, so that $X_e^{\rm tot}(z)$ approaches the PBH-modified residual history. At low redshift, the astrophysical contribution approaches $X_{\rm rei}(z)\to1+f_{\rm He}$. In what follows, we suppress the superscript ``tot'' when referring to the full ionization history. The precise low-redshift template does not affect our PBH bounds, which are derived exclusively from the interval $z\geq z_c=30$.

\subsection{Optical depth decomposition and high-redshift constraints}
\label{sec:tau}

The additional free electrons produced by PBH evaporation modify the redshift-dependent Thomson scattering rate and the CMB visibility function, leaving characteristic imprints on the temperature and polarization anisotropies. While the redshift dependence of the scattering determines the angular structure of these imprints, their cumulative amplitude can be conveniently characterized through the Thomson optical depth,
\begin{equation}
    \label{eq:opticaldepth}
    \tau(z) = \int_0^{z} {\rm d}z'\,
    \frac{c\,n_e(z')\,\sigma_T}{(1+z')H(z')}\,.
\end{equation}
Following Ref.~\cite{Cheng:2025cmb}, we decompose the total optical depth into low- and high-redshift contributions, $\tau = \tau_{\rm lowz}(z_c) + \tau_{\rm highz}(z_c)$. The transition redshift $z_c$ allows for a clean separation between the ionization contribution from known astrophysical reionization sources, such as stars and galaxies, which dominate at $z < z_c$, and that from exotic energy injection at higher redshifts. We focus exclusively on $\tau_{\rm highz}(z_c)$ to constrain the PBH parameter space. We set $z_c = 30$ and adopt a robust constraint derived from a GP reconstruction of the reionization history $X_e(z)$ using \textit{Planck} 2018 low-$\ell$ polarization data~\cite{Cheng:2025cmb}. Specifically, that analysis performs Bayesian inference using the low-$\ell$ EE-only SimAll likelihood from the \textit{Planck} 2018 data release~\cite{Planck:2018nkj,Planck:2019nip}, implemented in \texttt{Cobaya}~\cite{Torrado:2020dgo} and coupled to a modified version of \texttt{CLASS}~\cite{Blas:2011rf} supporting non-parametric $X_e(z)$ reconstructions up to $z_{\rm max}=800$. Standard cosmological (here, $\Lambda$CDM) parameters are fixed to their \textit{Planck} 2018 best-fit values so that the inferred $\tau_{\rm highz}$ posterior is determined primarily by the low-$\ell$ polarization data. The construction and validation of this posterior are presented in detail in Ref.~\cite{Cheng:2025cmb}, together with its public numerical implementation. In the present work, we use this posterior as an external statistical constraint on the PBH-induced contribution to $\tau_{\rm highz}$, following the same division between methodology and application adopted in Ref.~\cite{Yin:2025amn}.

We also restrict the PBH-induced energy injection to be post-recombination and within the redshift range covered by the GP reconstruction. We adopt $z_{\rm max}\simeq 800$ as a conservative upper limit: for injections approaching recombination, modifications to $X_e(z)$ primarily deform the visibility function around last scattering and become increasingly degenerate with recombination-era physics and standard cosmological parameters (such as $n_s$ or $Y_p$), limiting the robustness of constraints derived from low-$\ell$ polarization alone. We focus on the interval $30 \le z \le 800$. This redshift window defines the PBH mass interval
\begin{equation}
    \label{eq:massrange}
    3.2\times10^{13}\,{\rm g} \lesssim M_{\rm PBH} \lesssim 5\times10^{14}\,{\rm g}\,,
\end{equation}
corresponding to Hawking temperatures
\begin{equation}
    330\,{\rm MeV} \gtrsim T_H \gtrsim 21\,{\rm MeV}\,.
\end{equation}
The lower mass limit corresponds to PBHs that fully evaporate by $z_{\rm max}$, ensuring their energy injection is treated consistently as a perturbation to the optical depth rather than a modification to the recombination history. The upper limit corresponds to PBHs evaporating at the present day, which inject energy continuously throughout the post-recombination epoch. The EM cascades from these evaporating PBHs are modeled using \texttt{BlackHawk} and \texttt{DarkHistory}, accounting for the prompt hadronization and decay of primary particles.

\section{Results}
\label{sec:results}

Within the mass window defined in Eq.~\eqref{eq:massrange}, we assume a monochromatic PBH mass function and generate complete evaporation histories at representative mass points, spaced by $\Delta M_{\rm PBH} = 1.6 \times 10^{13}\,{\rm g}$ across the mass interval in Eq.~\eqref{eq:massrange}. Each mass point is treated independently, providing a full time-dependent emission history analogous to scanning over different particle physics scenarios for the injected spectra.

\begin{figure}[thb]
    \centering
    \includegraphics[width=\linewidth]{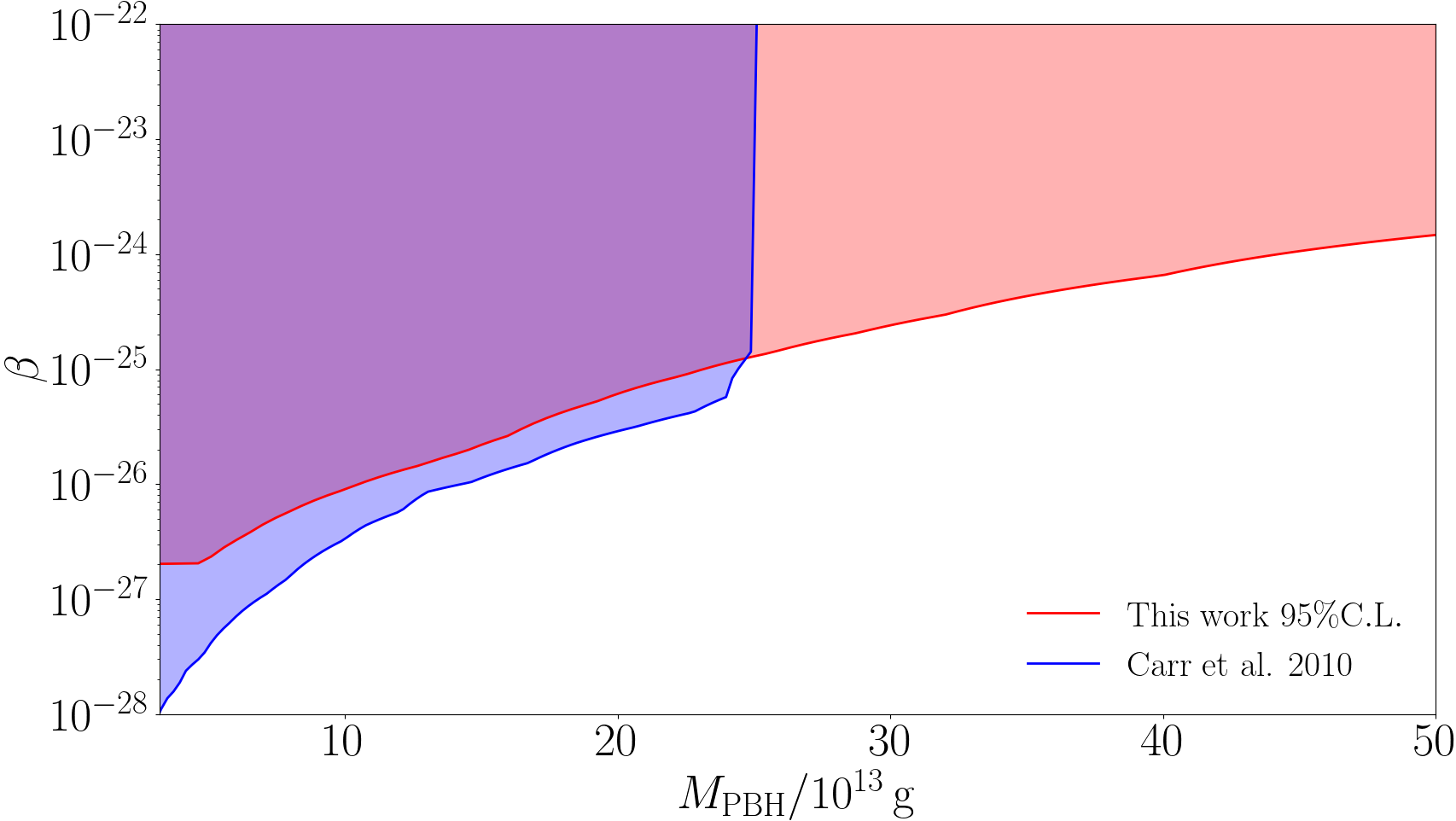}
    \caption{Upper bounds at 95\% C.L.\ on the initial PBH mass fraction at formation, $\beta$, as a function of the PBH mass for non-spinning black holes. The mass range shown corresponds to PBHs whose EM energy injection occurs predominantly at redshifts $z < z_{\rm max}$. The constraints are derived from the impact of Hawking evaporation on the thermal and ionization history of the IGM, as inferred from the high-redshift optical depth.}
    \label{fig:beta_constraints}
\end{figure}

Figure~\ref{fig:beta_constraints} shows the resulting constraints on the initial PBH mass fraction at formation, $\beta$, as a function of the PBH mass, obtained from the effects of Hawking evaporation on the thermal and ionization history of the Universe at 95\% confidence level (C.L.). The upper bound on $\beta$ is obtained by requiring the total predicted high-redshift optical depth to remain below the 95\% upper limit of the posterior shown in Fig.~2 of Ref.~\cite{Cheng:2025cmb}. The resulting constraint exhibits a mild nonlinear dependence on the PBH mass. Heavier PBHs inject energy at later times, so the resulting increase in the free-electron fraction occurs predominantly at lower redshifts. Because the optical-depth integral in Eq.~\eqref{eq:opticaldepth} is weighted toward higher redshift, $\tau(z) \propto \int_0^z {\rm d}z'\, X_e(z')\,(1+z')^{1/2}$, ionization occurring at late times is less efficient at enhancing $\tau(z)$, leading to weaker constraints. In addition, as the Hawking temperature increases toward lower PBH masses, the opening of hadronic and heavier SM emission channels redistributes the PBH luminosity among all species. After folding in hadronization, decays, and EM cascades from the PBH evolution and the particles propagation codes, this redistribution reduces the effective EM branching ratio, defined as the fraction of the total Hawking power that ultimately appears in photons and electrons/positrons and can efficiently couple to the IGM. This suppresses the impact of lighter PBHs on reionization observables and contributes to the weakening of the constraint at the low-mass end of the window.

\begin{figure}
    \centering
    \includegraphics[width=1\linewidth]{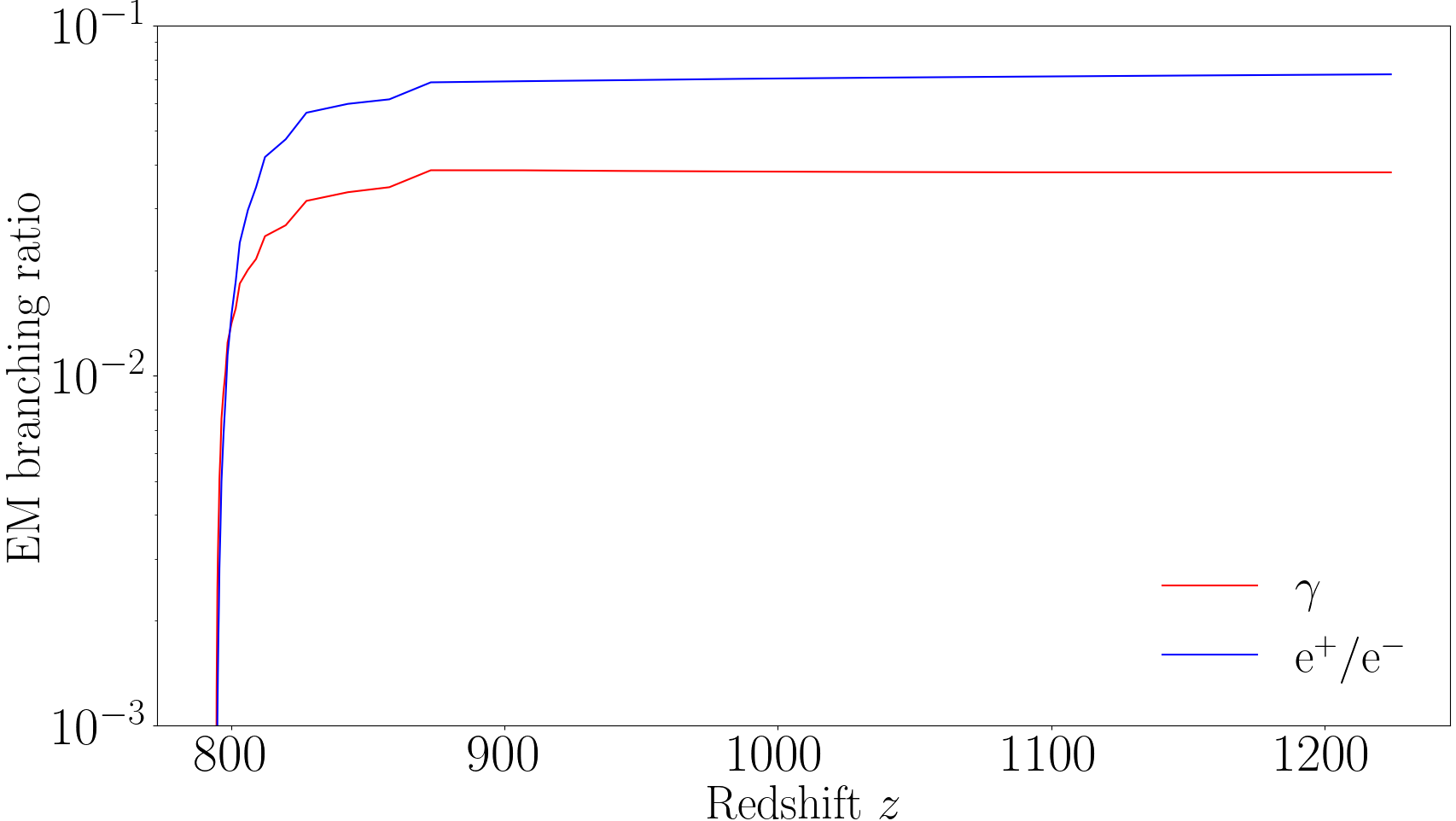}
    \caption{Time evolution of the effective EM branching ratio into photons (red) and electrons/positrons (blue) for PBH evaporation with mass $M_{\rm PBH} = 3.2\times 10^{13}\,\mathrm{g}$ and initial abundance $\beta = 1.6\times10^{-27}$. The EM branching ratio is defined as the fraction of the total Hawking luminosity that ultimately appears in electromagnetically interacting particles after hadronization, decays, and secondary EM cascades.}
    \label{fig:branch_ratio}
\end{figure}

Our results in Fig.~\ref{fig:beta_constraints} are compared with the CMB-based constraints derived in Ref.~\cite{Carr:2009jm}. In that work, bounds in the mass range $M_{\rm PBH} \sim 10^{13}\textrm{--}10^{14}\,\mathrm{g}$ are primarily driven by the damping of CMB temperature anisotropies due to energy injection around recombination, modeled using analytic estimates of the PBH lifetime and simplified prescriptions for energy deposition.\footnote{Other works have also derived CMB anisotropy constraints on light PBHs using increasingly detailed energy injection and recombination modeling, including comprehensive treatments of EM cascades and Boltzmann evolution~\cite{Poulin:2016anj, Stocker:2018avm, Poulter:2019ooo, Lucca:2019rxf}.} By contrast, the constraints presented here rely on a fully time-dependent numerical treatment of PBH evaporation and EM cascades, and are derived from the reconstructed high-redshift optical depth using a GP approach that is explicitly insensitive to late-time astrophysical reionization~\cite{Cheng:2025cmb}. This strategy allows us to isolate the contribution of exotic energy injection at $z \gtrsim 30$ and to probe PBH evaporation effects extending beyond the recombination epoch. As a result, our bounds provide a complementary constraint on the initial PBH abundance based on the improved sensitivity of modern high-redshift ionization measurements, providing a robust constraint complementary to earlier CMB limits. At the lowest PBH masses, our analysis yields more conservative limits than Ref.~\cite{Carr:2009jm}, reflecting both the restriction to low-$\ell$ polarization data and the fully time-dependent treatment of PBH evaporation and EM energy deposition. In particular, the effective EM branching ratio, obtained after consistently accounting for both the primary Hawking emission spectrum and the full secondary particle production, including hadronization and decays, is strongly mass- and redshift-dependent, as shown in Fig.~\ref{fig:branch_ratio} for a PBH of mass $3.2\times10^{13}\,{\rm g}$ evaporating at $z\simeq 800$. At low PBH masses, additional emission channels reduce the fraction of energy in photons and electrons, partially relaxing previous bounds that assumed a fixed EM fraction. The branching ratio evolution, combined with the full time-dependent emission treatment, explains the observed mild suppression of constraints in the low-mass regime of Fig.~\ref{fig:beta_constraints}.  

\section{Discussion}
\label{sec:discussion}

\begin{figure}
    \centering
    \includegraphics[width=1\linewidth]{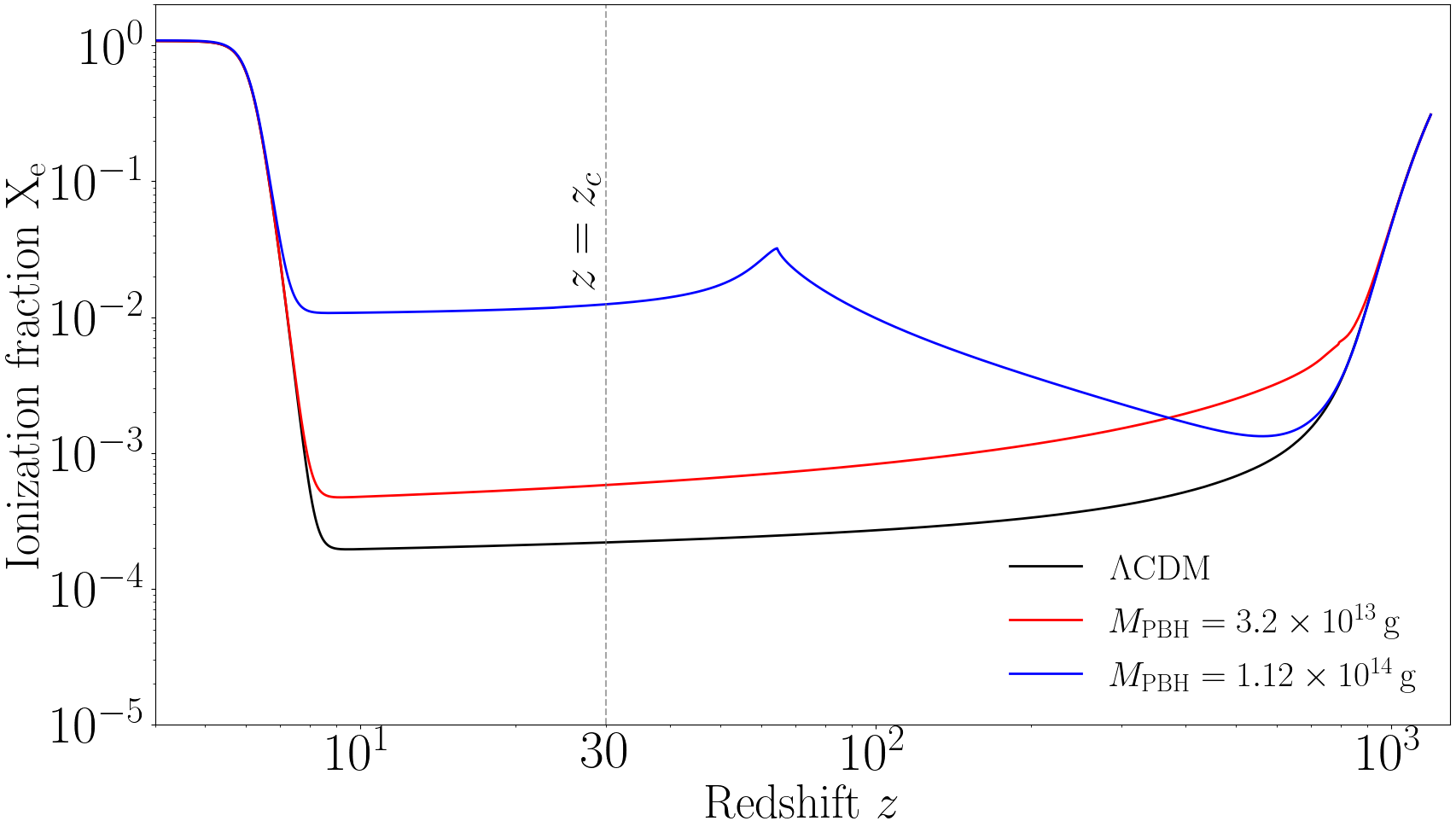}
    \caption{Hydrogen ionization fraction $X_e(z)$ as a function of redshift for two representative PBH populations evaluated near their respective upper limits on the initial PBH mass fraction $\beta$: $M_{\rm PBH}=3.2\times10^{13}\,\mathrm{g}$ (red) and $M_{\rm PBH}=1.12\times10^{14}\,\mathrm{g}$ (blue). The black solid line shows the baseline $\Lambda$CDM ionization history.}
    \label{fig:Xe}
\end{figure}

The time-dependent emission histories allow us to study the detailed impact of PBH evaporation on the IGM and baryon--photon plasma. Figure~\ref{fig:Xe} shows the evolution of the hydrogen ionization fraction, $X_e(z)$, for PBHs of different masses near their respective upper bounds on the initial PBH mass fraction $\beta$, compared to the baseline $\Lambda$CDM evolution. For PBH masses $M_{\rm PBH} \gtrsim 3.2\times 10^{13}\,\mathrm{g}$, evaporation occurs at redshifts $z < z_{\rm max}$, producing a characteristic spike in $X_e(z)$ at the epoch when most of the EM energy is injected into the IGM. In the figure and in what follows, we show two representative PBH populations: one with $M_{\rm PBH} = 3.2\times 10^{13}\,\mathrm{g}$ and $\beta = 1.6\times 10^{-27}$, corresponding to PBHs that evaporate close to the maximum redshift considered in our analysis, and a second with $M_{\rm PBH} = 1.12\times 10^{14}\,\mathrm{g}$ and $\beta = 1.4\times 10^{-26}$, representative of heavier PBHs that evaporate at later times. In both cases, the chosen values of $\beta$ lie close to the corresponding upper limits derived from the optical depth constraint, and are illustrative of the maximal ionization signatures consistent with current data.

The amplitude of the resulting spike in $X_e(z)$ is modulated by the EM branching ratio, which is computed numerically from the full Hawking emission spectrum and accounts for the suppression of EM energy at earlier evaporation times, when a larger fraction of the emitted power is carried by non-EM interacting particles. The enhancement of the ionization fraction is qualitatively consistent with earlier studies of post-recombination energy injection from DM annihilation~\cite{Cirelli:2009bb}, leading to a generic response of the IGM to exotic energy deposition. PBHs can substantially modify the ionization history of the Universe during the post-recombination epoch. While our analysis uses the high-redshift optical depth $\tau_{\rm highz}$ as a derived constraint, a direct likelihood evaluation of the PBH-induced ionization histories within the GP/CMB pipeline could exploit the detailed redshift dependence of the spike and potentially yield tighter constraints. By contrast, the current approach is conservative, as $\tau_{\rm highz}$ effectively integrates over the ionization history and does not fully exploit transient enhancements in $X_e$ at specific redshifts.

\begin{figure}
    \centering
    \includegraphics[width=1\linewidth]{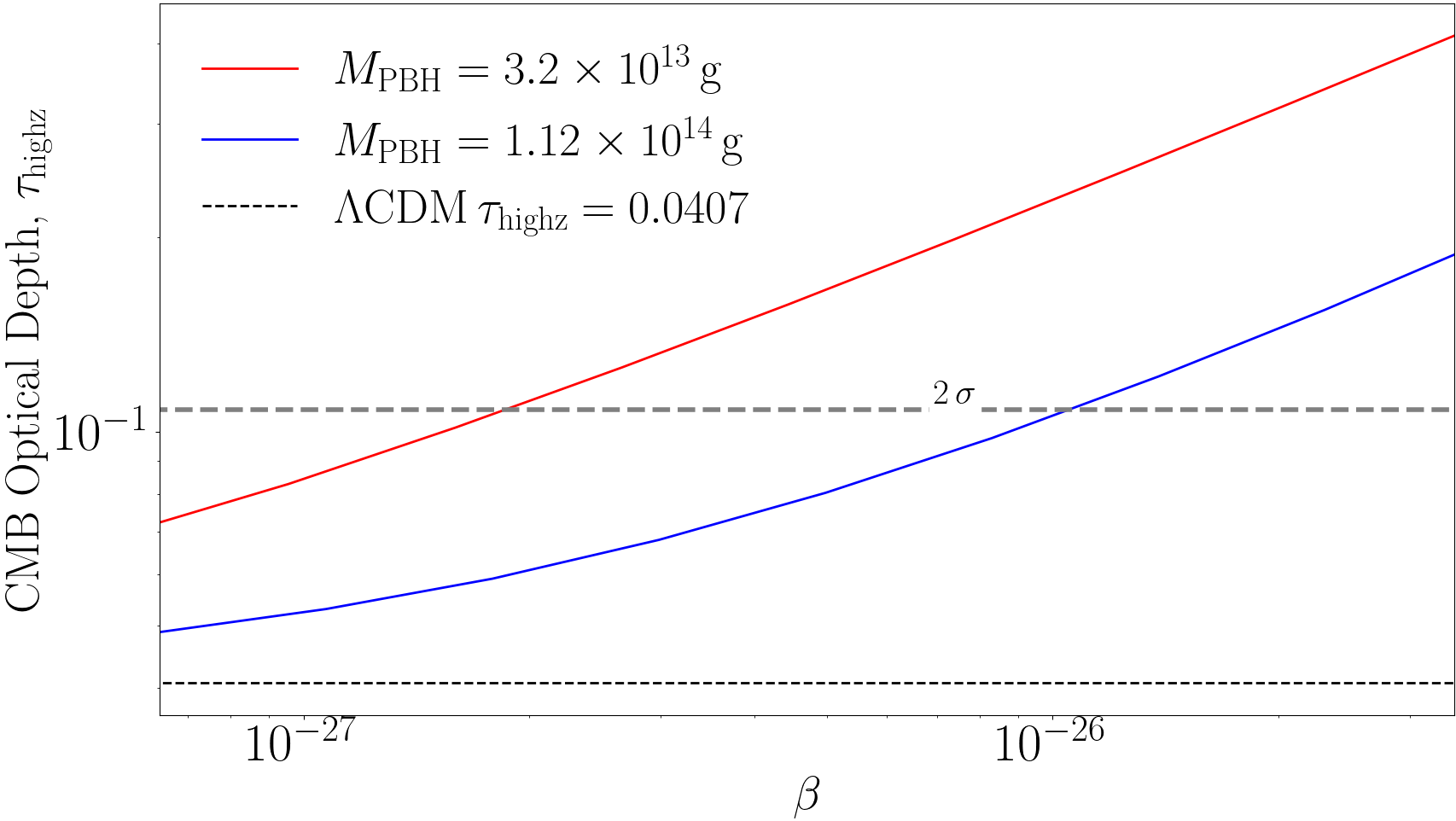}
    \caption{Total high-redshift CMB optical depth $\tau_{\rm highz}$ predicted in the presence of PBH evaporation as a function of $\beta$. The red solid line corresponds to PBHs with mass $M_{\rm PBH}=3.2\times10^{13}\,\mathrm{g}$, while the blue solid line corresponds to $M_{\rm PBH}=1.12\times10^{14}\,\mathrm{g}$. For each mass, $\tau_{\rm highz}$ is obtained by integrating the modified ionization history resulting from the full time-dependent Hawking evaporation and EM energy deposition into the IGM at redshifts above $z_c$. The black dashed line denotes the $\Lambda$CDM baseline contribution in the absence of PBHs. The horizontal gray dashed line indicates the 95\% C.L. upper limit on $\tau_{\rm highz}$ inferred from the GP reconstruction of the reionization history in Ref.~\cite{Cheng:2025cmb}. The intersection of this bound with each colored curve determines the corresponding upper limit on $\beta$.}
    \label{fig:tau_beta}
\end{figure}

We next consider two representative PBH populations with initial masses $M_{\rm PBH}^{\rm ini}=3.2\times10^{13},\mathrm{g}$ and $M_{\rm PBH}^{\rm ini}=1.12\times10^{14},\mathrm{g}$. Figure~\ref{fig:tau_beta} shows the total high-redshift optical depth as a function of the initial abundance $\beta$ for these two masses and illustrates how the GP constraint on $\tau_{\rm highz}$ translates into corresponding upper limits on $\beta$. As a reference point, for the lighter population, $\beta=1.6\times10^{-27}$ corresponds to an effective PBH dark-matter fraction $f_{\rm PBH}\simeq5\times10^{-9}$. For each value of $\beta$, the optical depth is obtained by integrating the modified ionization history $X_e(z)$ resulting from the time-dependent PBH evaporation history and the resulting EM energy deposition into the IGM. The increase in $\tau_{\rm highz}$ with $\beta$ reflects the fact that a larger initial PBH abundance produces a higher energy-injection rate and hence an enhanced free-electron fraction during the post-recombination epoch. The dependence on the PBH mass arises from both the redshift at which most of the evaporation power is released and the evolving EM branching ratio. For the two benchmark masses shown, the lighter PBHs evaporate earlier and inject a larger fraction of their energy at higher redshift, producing a more rapid increase in $\tau_{\rm highz}$ for a given $\beta$. The heavier PBHs evaporate more slowly, leading to a more extended and less efficient contribution to the high-redshift optical depth. The black dashed line indicates the $\Lambda$CDM baseline contribution to $\tau_{\rm highz}$ in the absence of PBHs, while the horizontal gray dashed line denotes the 95\% credible upper limit inferred from the GP reconstruction of the reionization history~\cite{Cheng:2025cmb}. Requiring the total predicted high-redshift optical depth to remain below this limit yields an upper bound on $\beta$, determined by the intersection of each curve with the gray dashed line. This procedure provides a direct mapping between the GP-derived reionization constraint and the PBH parameter space without imposing a fixed parametric model for late-time astrophysical reionization.

\begin{figure}
    \centering
    \includegraphics[width=1\linewidth]{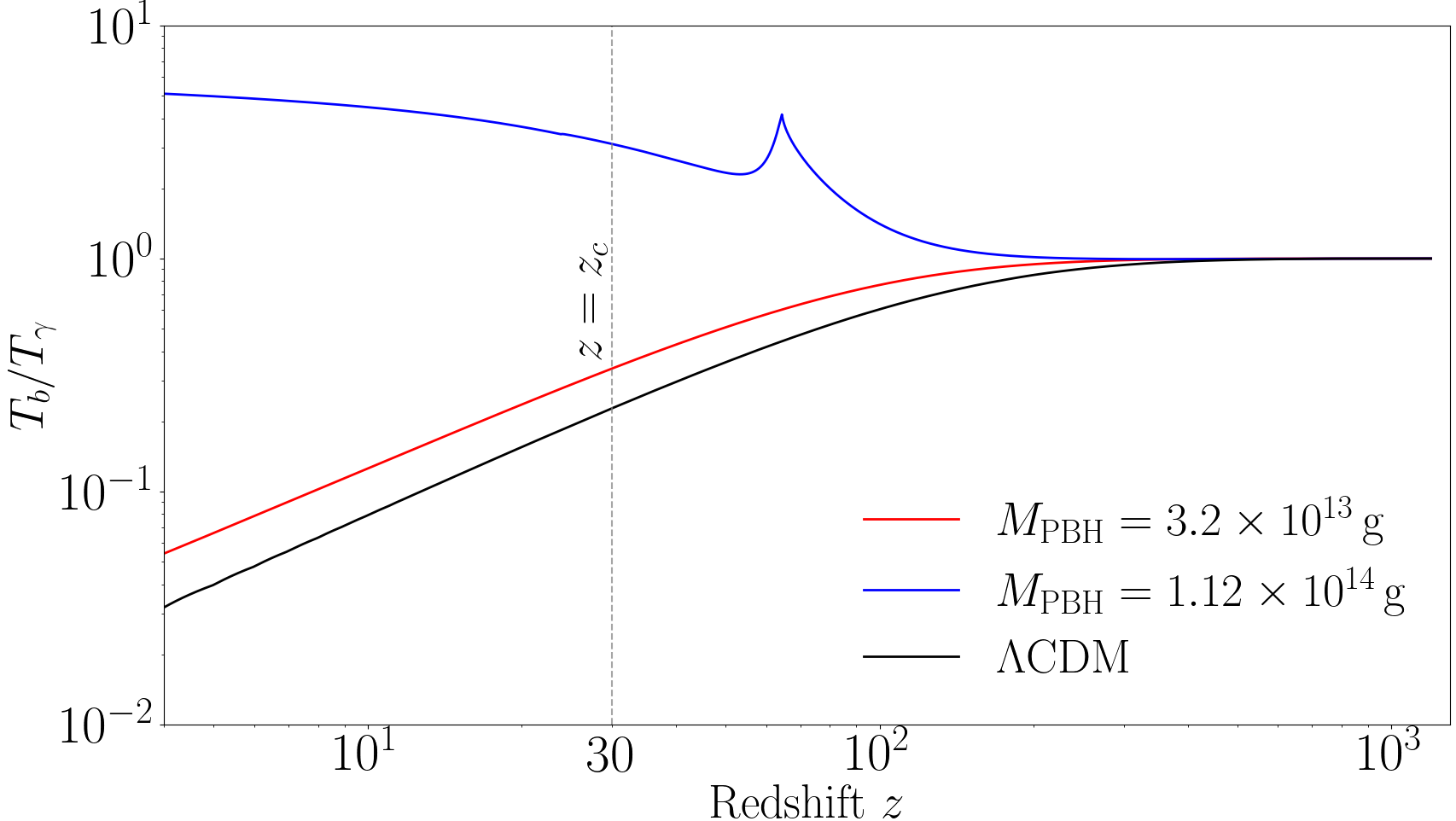}
    \caption{Ratio of the baryon temperature $T_b$ to the CMB photon temperature $T_\gamma$ as a function of redshift. The black line shows the standard $\Lambda$CDM prediction. Colored lines include the effect of PBH evaporation for populations with masses $M_{\rm PBH} = 3.2\times10^{13}\,\mathrm{g}$ (red) and $1.12\times10^{14}\,\mathrm{g}$ (blue), and corresponding initial abundances $\beta = 1.6\times10^{-27}$ and $1.4\times10^{-26}$.}
    \label{fig:Tb}
\end{figure}

The impact on the thermal history is shown in Fig.~\ref{fig:Tb}, where the baryon temperature $T_b$ is compared to the CMB photon temperature $T_\gamma$ for the standard $\Lambda$CDM scenario and for the representative PBH population described above. As discussed below Eq.~\eqref{eq:Tb}, baryons decouple from CMB photons already in the standard scenario. Energy injection from Hawking radiation raises the baryon temperature and, through the enhanced free-electron fraction, may also prolong residual Compton coupling, partially compensating adiabatic cooling. As a result, the ratio $T_b/T_\gamma$ is enhanced relative to the $\Lambda$CDM prediction, being systematically larger over a broad redshift range. This deviation gives a clear signature of exotic energy injection, distinct from the standard thermal history, with an amplitude that depends sensitively on the PBH mass and initial abundance. Such heating of the baryonic gas would also have direct implications for high-redshift 21-cm observables, so that future measurements of the IGM thermal state can be used jointly with reionization constraints.

Although incorporating small-scale temperature anisotropies (high-$\ell$ TT) would generally be expected to strengthen our constraints, we restrict our analysis to low-$\ell$ polarization data for robustness. PBH evaporation across the entire mass range considered here increases the total optical depth $\tau$, leading to a global suppression of the CMB acoustic peaks by the screening factor $\exp(-2\tau)$. In addition, PBHs towards the lighter end of the mass range considered inject a significant amount of energy at high redshifts, increasing the post-recombination ionization fraction $X_e(z)$. This broadens the last scattering surface and enhances Silk damping, resulting in a characteristic suppression of the high-$\ell$ tail of the power spectrum, see e.g.\ Refs.~\cite{Poulin:2016anj, Stocker:2018avm, Poulter:2019ooo, Yang:2019bkk}. As a consequence, including high-$\ell$ TT data would provide stricter limits across the entire PBH mass range, with particularly strong constraints for lighter PBHs, for which these damping effects are most pronounced. However, extracting these constraints from high-$\ell$ data requires a full cosmological parameter inference to account for parameter degeneracies. The global suppression of the acoustic peaks is highly degenerate with the primordial scalar amplitude $A_s$ through the combination $A_s\exp(-2\tau)$, while modifications to the damping tail can be degenerate with parameters governing standard recombination physics, such as the primordial helium abundance $Y_p$ or the electron mass $m_e$~\cite{Planck:2018vyg}. Disentangling the PBH signal from these standard cosmological parameters requires a comprehensive and careful modeling of the parameter space. To avoid these model dependencies and the risk of parameter confusion, we rely exclusively on the low-$\ell$ EE polarization data. This approach yields a more conservative constraint that is driven primarily by the reionization history reconstruction, accepting the trade-off of more conservative limits on the lightest PBHs in exchange for greater reliability and reduced dependence on extended cosmological parameter modeling.

A useful comparison for our reionization bounds is provided by constraints on PBH evaporation derived from CMB spectral distortions~\cite{Tashiro:2008sf, Chluba:2013dna, Chluba:2019nxa, Lucca:2019rxf, Acharya:2019xla, Acharya:2020jbv}. Spectral distortions are generated when EM energy is injected into the CMB after photon number-changing processes, such as double Compton scattering and bremsstrahlung, become inefficient, at redshifts $z \sim 10^4\textrm{--}10^6$. Energy injection during this epoch produces departures from a blackbody spectrum, parametrized by the chemical potential $\mu$ and the Compton $y$ parameter, which cannot be fully thermalized and are tightly constrained by COBE/FIRAS (Cosmic Background Explorer-Far InfraRed Absolute Spectrophotometer) observations~\cite{Fixsen:1996nj, 2002ApJ...581..817F}. By contrast, the analysis presented here focuses on the impact of PBH evaporation on the thermal and ionization history of the IGM at lower redshifts, $z \lesssim 800$, where injected energy primarily modifies the free-electron fraction and the baryon temperature rather than the CMB frequency spectrum. Although both approaches probe energy injection from PBH evaporation, they are sensitive to distinct physical observables: spectral-distortion measurements constrain early-time departures from a Planck CMB spectrum, while reionization constraints are driven by late-time changes in the ionization and thermal state of the gas. As a result, in the regime where PBH evaporation occurs predominantly at $z \sim 30\textrm{--}800$, the reionization bounds derived here provide significantly stronger constraints. Future CMB spectral-distortion missions such as the Primordial Inflation Explorer (PIXIE)~\cite{Kogut:2011xw} and the Polarized Radiation Imaging and Spectroscopy Mission (PRISM)~\cite{PRISM:2013fvg}, together with improved high-redshift CMB polarization~\cite{SimonsObservatory:2018koc, LiteBIRD:2022cnt, Sehgal:2019ewc, CMB-HD:2022bsz} and 21-cm measurements~\cite{Munshi:2023buw, HERA:2025ajm}, will provide complementary and substantially more sensitive probes of PBH evaporation across a wide range of redshifts and observables.

\section{Conclusions}
\label{sec:conclusions}

In this work we have derived constraints on the initial abundance of evaporating primordial black holes (PBHs) using the reionization history of the Universe as a cosmological probe. We focused on PBHs with masses in the range $3.2\times10^{13}\,\mathrm{g} \lesssim M_{\rm PBH} \lesssim 5\times10^{14}\,\mathrm{g}$, corresponding to BHs whose evaporation injects energy predominantly at redshifts below a benchmark value $z_{\rm max}=800$, chosen to ensure a clean separation from recombination physics. In this mass window, constraints from gamma-ray observations weaken, while the ionization and thermal histories of the intergalactic medium provide a sensitive and complementary probe of Hawking evaporation.

Our analysis introduces a state-of-the-art framework that combines a fully time-dependent treatment of PBH evaporation with a model-independent reconstruction of the reionization history. The Hawking emission spectra, including all Standard Model degrees of freedom and gravitons, are computed using \texttt{BlackHawk 2.3}, with final-state radiation modeled using \texttt{HAZMA}, while EM cascades and energy deposition into ionization, excitation, and heating are consistently tracked using \texttt{DarkHistory}. These effects are incorporated into a Gaussian Process reconstruction of the free-electron fraction based on low-$\ell$ CMB polarization data from the \textit{Planck} 2018 release, allowing us to constrain PBH evaporation through its contribution to the high-redshift optical depth, $\tau_{\rm highz}$, without relying on parametric assumptions about the reionization history. This approach allows for a well-defined separation between exotic energy injection at $z \gtrsim z_c \sim 30$, where the Universe is not expected to be fully reionized, and late-time astrophysical reionization. As a result, the derived bounds on the initial PBH abundance are robust against uncertainties in astrophysical modeling and directly trace the redshift-dependent impact of Hawking evaporation on the ionization history.

We find that reionization data place meaningful constraints on the initial PBH mass fraction $\beta$ across the considered mass range, with a characteristic mass dependence driven by the timing and composition of the evaporated energy. For lighter PBHs, evaporation occurs closer to recombination, where an increasing fraction of the luminosity is carried by non-EM particles, leading to weaker constraints than estimates based on constant branching ratios. For heavier PBHs, the emission extends to lower redshifts, where ionization contributes less efficiently to $\tau_{\rm highz}$, and larger initial abundances are allowed. Compared to earlier CMB anisotropy-based limits, our constraints are generally more conservative, reflecting the fully time-dependent and self-consistent treatment of evaporation, particle cascades, and energy deposition.

More broadly, this work shows the power of reionization as a precision probe of exotic energy injection in the early Universe when analyzed with flexible, data-driven techniques. Future improvements in large-scale CMB polarization measurements, as well as forthcoming 21-cm observations, will significantly enhance sensitivity to deviations in the high-redshift ionization history and further strengthen constraints on PBH evaporation. The framework developed here is readily applicable to a wide class of scenarios involving unstable relics or late-time energy injection, providing a robust and model-independent pathway for testing new physics with cosmological data.

\begin{acknowledgments}
We thank Doddy Marsh and Eleonora Di Valentino for helpful discussions and guidance on the implementation of the Gaussian Process reconstruction method and the handling of the data. LV also thanks Fabio Iocco for useful suggestions. We acknowledge support by the National Natural Science Foundation of China (NSFC) through the grant No.\ 12350610240 ``Astrophysical Axion Laboratories''. LV also acknowledges support by Istituto Nazionale di Fisica Nucleare (INFN) through the Commissione Scientifica Nazionale 4 (CSN4) Iniziativa Specifica ``Quantum Universe'' (QGSKY). LV thanks the Tsung-Dao Lee Institute for hospitality during the final stages of this work. This publication is based upon work from the COST Actions ``COSMIC WISPers'' (CA21106) and ``Addressing observational tensions in cosmology with systematics and fundamental physics (CosmoVerse)'' (CA21136), both supported by COST (European Cooperation in Science and Technology). This work made use of the open source software matplotlib~\cite{2007CSE.....9...90H}, numpy~\cite{2020Natur.585..357H}, and scipy~\cite{2020NatMe..17..261V}.
\end{acknowledgments}

\bibliographystyle{apsrev4-1}
\bibliography{references}

@article{2007CSE.....9...90H,
	adsurl = {https://ui.adsabs.harvard.edu/abs/2007CSE.....9...90H},
	author = {{Hunter}, John D.},
	doi = {10.1109/MCSE.2007.55},
	journal = {Computing in Science and Engineering},
	month = may,
	number = {3},
	pages = {90-95},
	title = {{Matplotlib: A 2D Graphics Environment}},
	volume = {9},
	year = 2007}

@article{Schiavone:2021imu,
    author = "Schiavone, Francesco and Montanino, Daniele and Mirizzi, Alessandro and Capozzi, Francesco",
    title = "{Axion-like particles from primordial black holes shining through the Universe}",
    eprint = "2107.03420",
    archivePrefix = "arXiv",
    primaryClass = "hep-ph",
    doi = "10.1088/1475-7516/2021/08/063",
    journal = "JCAP",
    volume = "08",
    pages = "063",
    year = "2021"
}

@article{Cirelli:2009bb,
    author = "Cirelli, Marco and Iocco, Fabio and Panci, Paolo",
    title = "{Constraints on Dark Matter annihilations from reionization and heating of the intergalactic gas}",
    eprint = "0907.0719",
    archivePrefix = "arXiv",
    primaryClass = "astro-ph.CO",
    reportNumber = "SACLAY-T09-065",
    doi = "10.1088/1475-7516/2009/10/009",
    journal = "JCAP",
    volume = "10",
    pages = "009",
    year = "2009"
}

@article{Vitale:2026ric,
    author = "Vitale, Giuseppe Filiberto and Lambiase, Gaetano and Poddar, Tanmay Kumar and Visinelli, Luca",
    title = "{Microscopic primordial black holes as macroscopic dark matter from large extra dimensions}",
    eprint = "2604.14871",
    archivePrefix = "arXiv",
    primaryClass = "astro-ph.CO",
    doi = "10.1088/1475-7516/2026/08/025",
    journal = "JCAP",
    volume = "08",
    pages = "025",
    year = "2026"
}

@article{Korwar:2023kpy,
    author = "Korwar, Mrunal and Profumo, Stefano",
    title = "{Updated constraints on primordial black hole evaporation}",
    eprint = "2302.04408",
    archivePrefix = "arXiv",
    primaryClass = "hep-ph",
    doi = "10.1088/1475-7516/2023/05/054",
    journal = "JCAP",
    volume = "05",
    pages = "054",
    year = "2023"
}

@article{Khan:2025kag,
    author = "Khan, Nabendu Kumar and Ray, Anupam and Kulkarni, Girish and Dasgupta, Basudeb",
    title = "{Stronger constraints on primordial black holes as dark matter derived from the thermal evolution of the intergalactic medium over the last twelve billion years}",
    eprint = "2503.15595",
    archivePrefix = "arXiv",
    primaryClass = "astro-ph.CO",
    reportNumber = "TIFR/TH/25-7, N3AS-25-004",
    doi = "10.1103/8h9x-hfpk",
    journal = "Phys. Rev. D",
    volume = "112",
    number = "12",
    pages = "123019",
    year = "2025"
}

@article{Ray:2021mxu,
    author = "Ray, Anupam and Laha, Ranjan and Mu{\~n}oz, Julian B. and Caputo, Regina",
    title = "{Near future MeV telescopes can discover asteroid-mass primordial black hole dark matter}",
    eprint = "2102.06714",
    archivePrefix = "arXiv",
    primaryClass = "astro-ph.CO",
    reportNumber = "TIFR/TH/21-1",
    doi = "10.1103/PhysRevD.104.023516",
    journal = "Phys. Rev. D",
    volume = "104",
    number = "2",
    pages = "023516",
    year = "2021"
}

@article{Wang:2025hwc,
    author = "Wang, Haonan and Zhang, Ying-li and Suyama, Teruaki",
    title = "{Nearly Monochromatic Primordial Black Holes as total Dark Matter from Bubble Collapse}",
    eprint = "2510.19233",
    archivePrefix = "arXiv",
    primaryClass = "astro-ph.CO",
    month = "10",
    year = "2025"
}

@article{Yang:2019bkk,
    author = "Yang, Yupeng",
    title = "{Constraints on the small scale curvature perturbation using Planck-2015 data}",
    eprint = "1904.09104",
    archivePrefix = "arXiv",
    primaryClass = "astro-ph.CO",
    doi = "10.1093/mnras/stz1148",
    journal = "Mon. Not. Roy. Astron. Soc.",
    volume = "486",
    number = "4",
    pages = "4569--4573",
    year = "2019"
}

@ARTICLE{2002ApJ...581..817F,
       author = {{Fixsen}, D.~J. and {Mather}, J.~C.},
        title = "{The Spectral Results of the Far-Infrared Absolute Spectrophotometer Instrument on COBE}",
      journal = {\apj},
         year = 2002,
        month = dec,
       volume = {581},
       number = {2},
        pages = {817-822},
          doi = {10.1086/344402},
       adsurl = {https://ui.adsabs.harvard.edu/abs/2002ApJ...581..817F}
}

@article{Harada:2017fjm,
    author = "Harada, Tomohiro and Yoo, Chul-Moon and Kohri, Kazunori and Nakao, Ken-Ichi",
    title = "{Spins of primordial black holes formed in the matter-dominated phase of the Universe}",
    eprint = "1707.03595",
    archivePrefix = "arXiv",
    primaryClass = "gr-qc",
    reportNumber = "KEK-COSMO-206, RUP-17-12, KEK-Cosmo-206, KEK-TH-1987, OCU-PHYS-466, AP-GR-139",
    doi = "10.1103/PhysRevD.96.083517",
    journal = "Phys. Rev. D",
    volume = "96",
    number = "8",
    pages = "083517",
    year = "2017",
    note = "[Erratum: Phys.Rev.D 99, 069904 (2019)]"
}

@article{DeLuca:2019buf,
    author = "De Luca, V. and Desjacques, V. and Franciolini, G. and Malhotra, A. and Riotto, A.",
    title = "{The initial spin probability distribution of primordial black holes}",
    eprint = "1903.01179",
    archivePrefix = "arXiv",
    primaryClass = "astro-ph.CO",
    doi = "10.1088/1475-7516/2019/05/018",
    journal = "JCAP",
    volume = "05",
    pages = "018",
    year = "2019"
}

@article{Lucca:2019rxf,
    author = {Lucca, Matteo and Sch{\"o}neberg, Nils and Hooper, Deanna C. and Lesgourgues, Julien and Chluba, Jens},
    title = "{The synergy between CMB spectral distortions and anisotropies}",
    eprint = "1910.04619",
    archivePrefix = "arXiv",
    primaryClass = "astro-ph.CO",
    reportNumber = "TTK-19-39, ULB-TH/19-08",
    doi = "10.1088/1475-7516/2020/02/026",
    journal = "JCAP",
    volume = "02",
    pages = "026",
    year = "2020"
}

@article{LiteBIRD:2022cnt,
    author = "Allys, E. and others",
    collaboration = "LiteBIRD",
    title = "{Probing Cosmic Inflation with the LiteBIRD Cosmic Microwave Background Polarization Survey}",
    eprint = "2202.02773",
    archivePrefix = "arXiv",
    primaryClass = "astro-ph.IM",
    doi = "10.1093/ptep/ptac150",
    journal = "PTEP",
    volume = "2023",
    number = "4",
    pages = "042F01",
    year = "2023"
}

@article{HERA:2025ajm,
    author = "Abdurashidova, Zuhra and others",
    collaboration = "HERA",
    title = "{First Results from HERA Phase II}",
    eprint = "2511.21289",
    archivePrefix = "arXiv",
    primaryClass = "astro-ph.CO",
    doi = "10.3847/1538-4357/ae2d54",
    journal = "Astrophys. J.",
    volume = "998",
    number = "1",
    pages = "33",
    year = "2026"
}

@article{Munshi:2023buw,
    author = "Munshi, S. and others",
    title = "{First upper limits on the 21 cm signal power spectrum from cosmic dawn from one night of observations with NenuFAR}",
    eprint = "2311.05364",
    archivePrefix = "arXiv",
    primaryClass = "astro-ph.CO",
    doi = "10.1051/0004-6361/202348329",
    journal = "Astron. Astrophys.",
    volume = "681",
    pages = "A62",
    year = "2024"
}

@article{CMB-HD:2022bsz,
    author = "Aiola, Simone and others",
    collaboration = "CMB-HD",
    title = "{Snowmass2021 CMB-HD White Paper}",
    eprint = "2203.05728",
    archivePrefix = "arXiv",
    primaryClass = "astro-ph.CO",
    reportNumber = "FERMILAB-PUB-22-344-PPD",
    month = "3",
    year = "2022"
}

@article{Sehgal:2019ewc,
    author = "Sehgal, Neelima and others",
    title = "{CMB-HD: An Ultra-Deep, High-Resolution Millimeter-Wave Survey Over Half the Sky}",
    eprint = "1906.10134",
    archivePrefix = "arXiv",
    primaryClass = "astro-ph.CO",
    journal = "Bull. Am. Astron. Soc.",
    volume = "51",
    number = "7",
    pages = "1--23",
    year = "2019"
}

@article{SimonsObservatory:2018koc,
    author = "Ade, Peter and others",
    collaboration = "Simons Observatory",
    title = "{The Simons Observatory: Science goals and forecasts}",
    eprint = "1808.07445",
    archivePrefix = "arXiv",
    primaryClass = "astro-ph.CO",
    doi = "10.1088/1475-7516/2019/02/056",
    journal = "JCAP",
    volume = "02",
    pages = "056",
    year = "2019"
}

@article{Serpico:2020ehh,
    author = "Serpico, Pasquale D. and Poulin, Vivian and Inman, Derek and Kohri, Kazunori",
    title = "{Cosmic microwave background bounds on primordial black holes including dark matter halo accretion}",
    eprint = "2002.10771",
    archivePrefix = "arXiv",
    primaryClass = "astro-ph.CO",
    reportNumber = "LAPTH-005/20, KEK-Cosmo-248, KEK-TH-2198, IPMU20-0021",
    doi = "10.1103/PhysRevResearch.2.023204",
    journal = "Phys. Rev. Res.",
    volume = "2",
    number = "2",
    pages = "023204",
    year = "2020"
}

@article{PRISM:2013fvg,
    author = "Andr{\'e}, Philippe and others",
    collaboration = "PRISM",
    title = "{PRISM (Polarized Radiation Imaging and Spectroscopy Mission): An Extended White Paper}",
    eprint = "1310.1554",
    archivePrefix = "arXiv",
    primaryClass = "astro-ph.CO",
    doi = "10.1088/1475-7516/2014/02/006",
    journal = "JCAP",
    volume = "02",
    pages = "006",
    year = "2014"
}

@article{Kogut:2011xw,
    author = "Kogut, A. and others",
    title = "{The Primordial Inflation Explorer (PIXIE): A Nulling Polarimeter for Cosmic Microwave Background Observations}",
    eprint = "1105.2044",
    archivePrefix = "arXiv",
    primaryClass = "astro-ph.CO",
    doi = "10.1088/1475-7516/2011/07/025",
    journal = "JCAP",
    volume = "07",
    pages = "025",
    year = "2011"
}

@article{2020Natur.585..357H,
	adsurl = {https://ui.adsabs.harvard.edu/abs/2020Natur.585..357H},
	archiveprefix = {arXiv},
	author = {{Harris}, Charles R. and others},
	doi = {10.1038/s41586-020-2649-2},
	eprint = {2006.10256},
	journal = {\nat},
	month = sep,
	number = {7825},
	pages = {357-362},
	primaryclass = {cs.MS},
	title = {{Array programming with NumPy}},
	volume = {585},
	year = 2020}

@article{2020NatMe..17..261V,
	adsurl = {https://ui.adsabs.harvard.edu/abs/2020NatMe..17..261V},
	archiveprefix = {arXiv},
	author = {{Virtanen}, Pauli and others},
	doi = {10.1038/s41592-019-0686-2},
	eprint = {1907.10121},
	journal = {Nature Methods},
	month = feb,
	pages = {261-272},
	primaryclass = {cs.MS},
	title = {{SciPy 1.0: fundamental algorithms for scientific computing in Python}},
	volume = {17},
	year = 2020}

@article{Yin:2025amn,
    author = "Yin, Ziwen and Cheng, Hanyu and Di Valentino, Eleonora and Gendler, Naomi and Marsh, David J. E. and Visinelli, Luca",
    title = "{Constraining the axiverse with reionization}",
    eprint = "2507.03535",
    archivePrefix = "arXiv",
    primaryClass = "hep-ph",
    reportNumber = "CA21106; CA21136",
    journal = {Phys. Rev. D},
    doi = "10.1103/wkm3-8qmh",
    month = "7",
    year = "2025"
}

@article{Lewis:2026mif,
    author = "Lewis, Antony",
    title = "{CAMB v2: cosmological power spectra for high-precision surveys}",
    eprint = "2607.14854",
    archivePrefix = "arXiv",
    primaryClass = "astro-ph.CO",
    month = "7",
    year = "2026"
}

@article{Lewis:2008wr,
    author = "Lewis, Antony",
    title = "{Cosmological parameters from WMAP 5-year temperature maps}",
    eprint = "0804.3865",
    archivePrefix = "arXiv",
    primaryClass = "astro-ph",
    doi = "10.1103/PhysRevD.78.023002",
    journal = "Phys. Rev. D",
    volume = "78",
    pages = "023002",
    year = "2008"
}

@article{Coogan:2019qpu,
    author = "Coogan, Adam and Morrison, Logan and Profumo, Stefano",
    title = "{Hazma: A Python Toolkit for Studying Indirect Detection of Sub-GeV Dark Matter}",
    eprint = "1907.11846",
    archivePrefix = "arXiv",
    primaryClass = "hep-ph",
    doi = "10.1088/1475-7516/2020/01/056",
    journal = "JCAP",
    volume = "01",
    pages = "056",
    year = "2020"
}

@article{Chen:2003gz,
    author = "Chen, Xue-Lei and Kamionkowski, Marc",
    title = "{Particle decays during the cosmic dark ages}",
    eprint = "astro-ph/0310473",
    archivePrefix = "arXiv",
    doi = "10.1103/PhysRevD.70.043502",
    journal = "Phys. Rev. D",
    volume = "70",
    pages = "043502",
    year = "2004"
}

@article{Padmanabhan:2005es,
    author = "Padmanabhan, Nikhil and Finkbeiner, Douglas P.",
    title = "{Detecting dark matter annihilation with CMB polarization: Signatures and experimental prospects}",
    eprint = "astro-ph/0503486",
    archivePrefix = "arXiv",
    doi = "10.1103/PhysRevD.72.023508",
    journal = "Phys. Rev. D",
    volume = "72",
    pages = "023508",
    year = "2005"
}

@article{Slatyer:2009yq,
    author = "Slatyer, Tracy R. and Padmanabhan, Nikhil and Finkbeiner, Douglas P.",
    title = "{CMB Constraints on WIMP Annihilation: Energy Absorption During the Recombination Epoch}",
    eprint = "0906.1197",
    archivePrefix = "arXiv",
    primaryClass = "astro-ph.CO",
    doi = "10.1103/PhysRevD.80.043526",
    journal = "Phys. Rev. D",
    volume = "80",
    pages = "043526",
    year = "2009"
}

@article{Slatyer:2015jla,
    author = "Slatyer, Tracy R.",
    title = "{Indirect dark matter signatures in the cosmic dark ages. I. Generalizing the bound on s-wave dark matter annihilation from Planck results}",
    eprint = "1506.03811",
    archivePrefix = "arXiv",
    primaryClass = "hep-ph",
    reportNumber = "MIT-CTP-4682",
    doi = "10.1103/PhysRevD.93.023527",
    journal = "Phys. Rev. D",
    volume = "93",
    number = "2",
    pages = "023527",
    year = "2016"
}

@article{Chluba:2013dna,
    author = "Chluba, Jens and Grin, Daniel",
    title = "{CMB spectral distortions from small-scale isocurvature fluctuations}",
    eprint = "1304.4596",
    archivePrefix = "arXiv",
    primaryClass = "astro-ph.CO",
    doi = "10.1093/mnras/stt1129",
    journal = "Mon. Not. Roy. Astron. Soc.",
    volume = "434",
    pages = "1619--1635",
    year = "2013"
}

@article{Clark:2018ghm,
    author = "Clark, Steven and Dutta, Bhaskar and Gao, Yu and Ma, Yin-Zhe and Strigari, Louis E.",
    title = "{21 cm limits on decaying dark matter and primordial black holes}",
    eprint = "1803.09390",
    archivePrefix = "arXiv",
    primaryClass = "astro-ph.HE",
    reportNumber = "MI-TH-1879",
    doi = "10.1103/PhysRevD.98.043006",
    journal = "Phys. Rev. D",
    volume = "98",
    number = "4",
    pages = "043006",
    year = "2018"
}

@article{Arbey:2019mbc,
    author = "Arbey, Alexandre and Auffinger, J{\'e}r{\'e}my",
    title = "{BlackHawk: A public code for calculating the Hawking evaporation spectra of any black hole distribution}",
    eprint = "1905.04268",
    archivePrefix = "arXiv",
    primaryClass = "gr-qc",
    reportNumber = "CERN-TH-2019-067",
    doi = "10.1140/epjc/s10052-019-7161-1",
    journal = "Eur. Phys. J. C",
    volume = "79",
    number = "8",
    pages = "693",
    year = "2019"
}

@article{Auffinger:2022sqj,
    author = "Auffinger, J{\'e}r{\'e}my and Arbey, Alexandre",
    title = "{Beyond the Standard Model with BlackHawk v2.0}",
    eprint = "2207.03266",
    archivePrefix = "arXiv",
    primaryClass = "gr-qc",
    doi = "10.22323/1.409.0017",
    journal = "PoS",
    volume = "CompTools2021",
    pages = "017",
    year = "2022"
}

@article{Mirocha:2020slz,
    author = "Mirocha, Jordan and La Plante, Paul and Liu, Adrian",
    title = "{The importance of galaxy formation histories in models of reionization}",
    eprint = "2012.09189",
    archivePrefix = "arXiv",
    primaryClass = "astro-ph.CO",
    doi = "10.1093/mnras/stab1871",
    journal = "Mon. Not. Roy. Astron. Soc.",
    volume = "507",
    number = "3",
    pages = "3872--3887",
    year = "2021"
}

@article{Chluba:2019nxa,
    author = "Chluba, J. and others",
    title = "{New horizons in cosmology with spectral distortions of the cosmic microwave background}",
    eprint = "1909.01593",
    archivePrefix = "arXiv",
    primaryClass = "astro-ph.CO",
    doi = "10.1007/s10686-021-09729-5",
    journal = "Exper. Astron.",
    volume = "51",
    number = "3",
    pages = "1515--1554",
    year = "2021"
}

@article{Cheng:2025cmb,
    author = "Cheng, Hanyu and Yin, Ziwen and Di Valentino, Eleonora and Marsh, David J. E. and Visinelli, Luca",
    title = "{Constraining exotic high-$z$ reionization histories with Gaussian processes and the Cosmic Microwave Background}",
    eprint = "2506.19096",
    archivePrefix = "arXiv",
    primaryClass = "astro-ph.CO",
    reportNumber = "CA21106; CA21136",
    month = "6",
    year = "2025"
}

@article{Carr:2009jm,
    author = "Carr, B. J. and Kohri, Kazunori and Sendouda, Yuuiti and Yokoyama, Jun'ichi",
    title = "{New cosmological constraints on primordial black holes}",
    eprint = "0912.5297",
    archivePrefix = "arXiv",
    primaryClass = "astro-ph.CO",
    reportNumber = "RESCEU-31-09, TU-852, YITP-09-112",
    doi = "10.1103/PhysRevD.81.104019",
    journal = "Phys. Rev. D",
    volume = "81",
    pages = "104019",
    year = "2010"
}

@article{Poulter:2019ooo,
    author = {Poulter, Harry and Ali-Ha{\"\i}moud, Yacine and Hamann, Jan and White, Martin and Williams, Anthony G.},
    title = "{CMB constraints on ultra-light primordial black holes with extended mass distributions}",
    eprint = "1907.06485",
    archivePrefix = "arXiv",
    primaryClass = "astro-ph.CO",
    month = "7",
    year = "2019"
}

@article{Stocker:2018avm,
    author = {St{\"o}cker, Patrick and Kr{\"a}mer, Michael and Lesgourgues, Julien and Poulin, Vivian},
    title = "{Exotic energy injection with ExoCLASS: Application to the Higgs portal model and evaporating black holes}",
    eprint = "1801.01871",
    archivePrefix = "arXiv",
    primaryClass = "astro-ph.CO",
    doi = "10.1088/1475-7516/2018/03/018",
    journal = "JCAP",
    volume = "03",
    pages = "018",
    year = "2018"
}

@article{Blas:2011rf,
    author = "Blas, Diego and Lesgourgues, Julien and Tram, Thomas",
    title = "{The Cosmic Linear Anisotropy Solving System (CLASS) II: Approximation schemes}",
    eprint = "1104.2933",
    archivePrefix = "arXiv",
    primaryClass = "astro-ph.CO",
    reportNumber = "CERN-PH-TH-2011-082, LAPTH-010-11",
    doi = "10.1088/1475-7516/2011/07/034",
    journal = "JCAP",
    volume = "07",
    pages = "034",
    year = "2011"
}

@article{Torrado:2020dgo,
    author = "Torrado, Jesus and Lewis, Antony",
    title = "{Cobaya: Code for Bayesian Analysis of hierarchical physical models}",
    eprint = "2005.05290",
    archivePrefix = "arXiv",
    primaryClass = "astro-ph.IM",
    reportNumber = "TTK-20-15",
    doi = "10.1088/1475-7516/2021/05/057",
    journal = "JCAP",
    volume = "05",
    pages = "057",
    year = "2021"
}

@article{Planck:2018nkj,
    author = "Aghanim, N. and others",
    collaboration = "Planck",
    title = "{Planck 2018 results. I. Overview and the cosmological legacy of Planck}",
    eprint = "1807.06205",
    archivePrefix = "arXiv",
    primaryClass = "astro-ph.CO",
    doi = "10.1051/0004-6361/201833880",
    journal = "Astron. Astrophys.",
    volume = "641",
    pages = "A1",
    year = "2020"
}

@article{Planck:2018vyg,
    author = "Aghanim, N. and others",
    collaboration = "Planck",
    title = "{Planck 2018 results. VI. Cosmological parameters}",
    eprint = "1807.06209",
    archivePrefix = "arXiv",
    primaryClass = "astro-ph.CO",
    doi = "10.1051/0004-6361/201833910",
    journal = "Astron. Astrophys.",
    volume = "641",
    pages = "A6",
    year = "2020",
    note = "[Erratum: Astron.Astrophys. 652, C4 (2021)]"
}

@article{Planck:2019nip,
    author = "Aghanim, N. and others",
    collaboration = "Planck",
    title = "{Planck 2018 results. V. CMB power spectra and likelihoods}",
    eprint = "1907.12875",
    archivePrefix = "arXiv",
    primaryClass = "astro-ph.CO",
    doi = "10.1051/0004-6361/201936386",
    journal = "Astron. Astrophys.",
    volume = "641",
    pages = "A5",
    year = "2020"
}

@article{Ghez:1998ph,
    author = "Ghez, A. M. and Klein, B. L. and Morris, M. and Becklin, E. E.",
    title = "{High proper motion stars in the vicinity of Sgr A*: Evidence for a supermassive black hole at the center of our galaxy}",
    eprint = "astro-ph/9807210",
    archivePrefix = "arXiv",
    doi = "10.1086/306528",
    journal = "Astrophys. J.",
    volume = "509",
    pages = "678--686",
    year = "1998"
}

@article{Poulin:2016anj,
    author = "Poulin, Vivian and Lesgourgues, Julien and Serpico, Pasquale D.",
    title = "{Cosmological constraints on exotic injection of electromagnetic energy}",
    eprint = "1610.10051",
    archivePrefix = "arXiv",
    primaryClass = "astro-ph.CO",
    doi = "10.1088/1475-7516/2017/03/043",
    journal = "JCAP",
    volume = "03",
    pages = "043",
    year = "2017"
}

@article{Mazde:2022sdx,
    author = "Mazde, Kratika and Visinelli, Luca",
    title = "{The interplay between the dark matter axion and primordial black holes}",
    eprint = "2209.14307",
    archivePrefix = "arXiv",
    primaryClass = "astro-ph.CO",
    doi = "10.1088/1475-7516/2023/01/021",
    journal = "JCAP",
    volume = "01",
    pages = "021",
    year = "2023"
}

@article{Ghez:2008ms,
    author = "Ghez, A. M. and others",
    title = "{Measuring Distance and Properties of the Milky Way's Central Supermassive Black Hole with Stellar Orbits}",
    eprint = "0808.2870",
    archivePrefix = "arXiv",
    primaryClass = "astro-ph",
    doi = "10.1086/592738",
    journal = "Astrophys. J.",
    volume = "689",
    pages = "1044--1062",
    year = "2008"
}

@article{Gillessen:2008qv,
    author = "Gillessen, S. and Eisenhauer, F. and Trippe, S. and Alexander, T. and Genzel, R. and Martins, F. and Ott, T.",
    title = "{Monitoring stellar orbits around the Massive Black Hole in the Galactic Center}",
    eprint = "0810.4674",
    archivePrefix = "arXiv",
    primaryClass = "astro-ph",
    doi = "10.1088/0004-637X/692/2/1075",
    journal = "Astrophys. J.",
    volume = "692",
    pages = "1075--1109",
    year = "2009"
}

@article{LIGOScientific:2016aoc,
    author = "Abbott, B. P. and others",
    collaboration = "LIGO Scientific, Virgo",
    title = "{Observation of Gravitational Waves from a Binary Black Hole Merger}",
    eprint = "1602.03837",
    archivePrefix = "arXiv",
    primaryClass = "gr-qc",
    reportNumber = "LIGO-P150914",
    doi = "10.1103/PhysRevLett.116.061102",
    journal = "Phys. Rev. Lett.",
    volume = "116",
    number = "6",
    pages = "061102",
    year = "2016"
}

@article{LIGOScientific:2018mvr,
    author = "Abbott, B. P. and others",
    collaboration = "LIGO Scientific, Virgo",
    title = "{GWTC-1: A Gravitational-Wave Transient Catalog of Compact Binary Mergers Observed by LIGO and Virgo during the First and Second Observing Runs}",
    eprint = "1811.12907",
    archivePrefix = "arXiv",
    primaryClass = "astro-ph.HE",
    reportNumber = "LIGO-P1800307",
    doi = "10.1103/PhysRevX.9.031040",
    journal = "Phys. Rev. X",
    volume = "9",
    number = "3",
    pages = "031040",
    year = "2019"
}

@article{EventHorizonTelescope:2019dse,
    author = "Akiyama, Kazunori and others",
    collaboration = "Event Horizon Telescope",
    title = "{First M87 Event Horizon Telescope Results. I. The Shadow of the Supermassive Black Hole}",
    eprint = "1906.11238",
    archivePrefix = "arXiv",
    primaryClass = "astro-ph.GA",
    doi = "10.3847/2041-8213/ab0ec7",
    journal = "Astrophys. J. Lett.",
    volume = "875",
    pages = "L1",
    year = "2019"
}

@article{EventHorizonTelescope:2022xnr,
    author = "Akiyama, Kazunori and others",
    collaboration = "Event Horizon Telescope",
    title = "{First Sagittarius A* Event Horizon Telescope Results. I. The Shadow of the Supermassive Black Hole in the Center of the Milky Way}",
    eprint = "2311.08680",
    archivePrefix = "arXiv",
    primaryClass = "astro-ph.HE",
    doi = "10.3847/2041-8213/ac6674",
    journal = "Astrophys. J. Lett.",
    volume = "930",
    number = "2",
    pages = "L12",
    year = "2022"
}

@article{Heger:2002by,
    author = "Heger, Alexander and Fryer, C. L. and Woosley, S. E. and Langer, N. and Hartmann, D. H.",
    title = "{How massive single stars end their life}",
    eprint = "astro-ph/0212469",
    archivePrefix = "arXiv",
    doi = "10.1086/375341",
    journal = "Astrophys. J.",
    volume = "591",
    pages = "288--300",
    year = "2003"
}

@article{Carr:1974nx,
    author = "Carr, Bernard J. and Hawking, S. W.",
    title = "{Black holes in the early Universe}",
    doi = "10.1093/mnras/168.2.399",
    journal = "Mon. Not. Roy. Astron. Soc.",
    volume = "168",
    pages = "399--415",
    year = "1974"
}

@article{Carr:1975qj,
    author = "Carr, Bernard J.",
    title = "{The Primordial black hole mass spectrum}",
    doi = "10.1086/153853",
    journal = "Astrophys. J.",
    volume = "201",
    pages = "1--19",
    year = "1975"
}

@article{Zeldovich:1967lct,
    author = "Zel'dovich, Ya. B. and Novikov, I. D.",
    title = "{The Hypothesis of Cores Retarded during Expansion and the Hot Cosmological Model}",
    journal = "Sov. Astron.",
    volume = "10",
    pages = "602",
    year = "1967"
}

@article{Shibata:1999zs,
    author = "Shibata, Masaru and Sasaki, Misao",
    title = "{Black hole formation in the Friedmann universe: Formulation and computation in numerical relativity}",
    eprint = "gr-qc/9905064",
    archivePrefix = "arXiv",
    reportNumber = "OU-TAP-93",
    doi = "10.1103/PhysRevD.60.084002",
    journal = "Phys. Rev. D",
    volume = "60",
    pages = "084002",
    year = "1999"
}

@article{Niemeyer:1999ak,
    author = "Niemeyer, Jens C. and Jedamzik, K.",
    title = "{Dynamics of primordial black hole formation}",
    eprint = "astro-ph/9901292",
    archivePrefix = "arXiv",
    doi = "10.1103/PhysRevD.59.124013",
    journal = "Phys. Rev. D",
    volume = "59",
    pages = "124013",
    year = "1999"
}

@article{Musco:2004ak,
    author = "Musco, Ilia and Miller, John C. and Rezzolla, Luciano",
    title = "{Computations of primordial black hole formation}",
    eprint = "gr-qc/0412063",
    archivePrefix = "arXiv",
    doi = "10.1088/0264-9381/22/7/013",
    journal = "Class. Quant. Grav.",
    volume = "22",
    pages = "1405--1424",
    year = "2005"
}

@article{Wu:2020ilx,
    author = "Wu, Yi-Peng",
    title = "{Peak statistics for the primordial black hole abundance}",
    eprint = "2005.00441",
    archivePrefix = "arXiv",
    primaryClass = "astro-ph.CO",
    doi = "10.1016/j.dark.2020.100654",
    journal = "Phys. Dark Univ.",
    volume = "30",
    pages = "100654",
    year = "2020"
}

@article{Gow:2020bzo,
    author = "Gow, Andrew D. and Byrnes, Christian T. and Cole, Philippa S. and Young, Sam",
    title = "{The power spectrum on small scales: Robust constraints and comparing PBH methodologies}",
    eprint = "2008.03289",
    archivePrefix = "arXiv",
    primaryClass = "astro-ph.CO",
    doi = "10.1088/1475-7516/2021/02/002",
    journal = "JCAP",
    volume = "02",
    pages = "002",
    year = "2021"
}

@article{Garcia-Bellido:1996mdl,
    author = "Garcia-Bellido, Juan and Linde, Andrei D. and Wands, David",
    title = "{Density perturbations and black hole formation in hybrid inflation}",
    eprint = "astro-ph/9605094",
    archivePrefix = "arXiv",
    reportNumber = "SU-ITP-96-20, SUSSEX-AST-96-5-1",
    doi = "10.1103/PhysRevD.54.6040",
    journal = "Phys. Rev. D",
    volume = "54",
    pages = "6040--6058",
    year = "1996"
}

@article{Flores:2020drq,
    author = "Flores, Marcos M. and Kusenko, Alexander",
    title = "{Primordial Black Holes from Long-Range Scalar Forces and Scalar Radiative Cooling}",
    eprint = "2008.12456",
    archivePrefix = "arXiv",
    primaryClass = "astro-ph.CO",
    reportNumber = "IPMU20-0092",
    doi = "10.1103/PhysRevLett.126.041101",
    journal = "Phys. Rev. Lett.",
    volume = "126",
    number = "4",
    pages = "041101",
    year = "2021"
}

@article{Maeso:2021xvl,
    author = {Maeso, David Navidad and Marzola, Luca and Raidal, Martti and Vaskonen, Ville and Veerm{\"a}e, Hardi},
    title = "{Primordial black holes from spectator field bubbles}",
    eprint = "2112.01505",
    archivePrefix = "arXiv",
    primaryClass = "astro-ph.CO",
    doi = "10.1088/1475-7516/2022/02/017",
    journal = "JCAP",
    volume = "02",
    number = "02",
    pages = "017",
    year = "2022"
}

@article{Byrnes:2018clq,
    author = "Byrnes, Christian T. and Hindmarsh, Mark and Young, Sam and Hawkins, Michael R. S.",
    title = "{Primordial black holes with an accurate QCD equation of state}",
    eprint = "1801.06138",
    archivePrefix = "arXiv",
    primaryClass = "astro-ph.CO",
    doi = "10.1088/1475-7516/2018/08/041",
    journal = "JCAP",
    volume = "08",
    pages = "041",
    year = "2018"
}

@article{Carr:2020gox,
    author = "Carr, Bernard and Kohri, Kazunori and Sendouda, Yuuiti and Yokoyama, Jun'ichi",
    title = "{Constraints on primordial black holes}",
    eprint = "2002.12778",
    archivePrefix = "arXiv",
    primaryClass = "astro-ph.CO",
    reportNumber = "RESCEU-03/20; KEK-Cosmo-249; KEK-TH-2199; IPMU20-0024",
    doi = "10.1088/1361-6633/ac1e31",
    journal = "Rept. Prog. Phys.",
    volume = "84",
    number = "11",
    pages = "116902",
    year = "2021"
}

@article{Carr:2020xqk,
    author = "Carr, Bernard and Kuhnel, Florian",
    title = "{Primordial Black Holes as Dark Matter: Recent Developments}",
    eprint = "2006.02838",
    archivePrefix = "arXiv",
    primaryClass = "astro-ph.CO",
    doi = "10.1146/annurev-nucl-050520-125911",
    journal = "Ann. Rev. Nucl. Part. Sci.",
    volume = "70",
    pages = "355--394",
    year = "2020"
}

@article{Hawking:1974rv,
    author = "Hawking, S. W.",
    title = "{Black hole explosions}",
    doi = "10.1038/248030a0",
    journal = "Nature",
    volume = "248",
    pages = "30--31",
    year = "1974"
}

@article{Ali-Haimoud:2016mbv,
    author = {Ali-Ha{\"\i}moud, Yacine and Kamionkowski, Marc},
    title = "{Cosmic microwave background limits on accreting primordial black holes}",
    eprint = "1612.05644",
    archivePrefix = "arXiv",
    primaryClass = "astro-ph.CO",
    doi = "10.1103/PhysRevD.95.043534",
    journal = "Phys. Rev. D",
    volume = "95",
    number = "4",
    pages = "043534",
    year = "2017"
}

@article{Fixsen:1996nj,
    author = "Fixsen, D. J. and Cheng, E. S. and Gales, J. M. and Mather, John C. and Shafer, R. A. and Wright, E. L.",
    title = "{The Cosmic Microwave Background spectrum from the full COBE FIRAS data set}",
    eprint = "astro-ph/9605054",
    archivePrefix = "arXiv",
    doi = "10.1086/178173",
    journal = "Astrophys. J.",
    volume = "473",
    pages = "576",
    year = "1996"
}

@article{Acharya:2020jbv,
    author = "Acharya, Sandeep Kumar and Khatri, Rishi",
    title = "{CMB and BBN constraints on evaporating primordial black holes revisited}",
    eprint = "2002.00898",
    archivePrefix = "arXiv",
    primaryClass = "astro-ph.CO",
    doi = "10.1088/1475-7516/2020/06/018",
    journal = "JCAP",
    volume = "06",
    pages = "018",
    year = "2020"
}

@article{Niikura:2017zjd,
    author = "Niikura, Hiroko and others",
    title = "{Microlensing constraints on primordial black holes with Subaru/HSC Andromeda observations}",
    eprint = "1701.02151",
    archivePrefix = "arXiv",
    primaryClass = "astro-ph.CO",
    doi = "10.1038/s41550-019-0723-1",
    journal = "Nature Astron.",
    volume = "3",
    number = "6",
    pages = "524--534",
    year = "2019"
}

@article{Carr:1997cn,
    author = "Carr, Bernard J. and Sakellariadou, M.",
    title = "{Dynamical constraints on dark compact objects}",
    reportNumber = "FERMILAB-PUB-97-299-A",
    doi = "10.1086/307071",
    journal = "Astrophys. J.",
    volume = "516",
    pages = "195--220",
    year = "1999"
}

@article{Carr:2020erq,
    author = "Carr, Bernard and Kuhnel, Florian and Visinelli, Luca",
    title = "{Constraints on Stupendously Large Black Holes}",
    eprint = "2008.08077",
    archivePrefix = "arXiv",
    primaryClass = "astro-ph.CO",
    doi = "10.1093/mnras/staa3651",
    journal = "Mon. Not. Roy. Astron. Soc.",
    volume = "501",
    number = "2",
    pages = "2029--2043",
    year = "2021"
}

@article{Coogan:2020tuf,
    author = "Coogan, Adam and Morrison, Logan and Profumo, Stefano",
    title = "{Direct Detection of Hawking Radiation from Asteroid-Mass Primordial Black Holes}",
    eprint = "2010.04797",
    archivePrefix = "arXiv",
    primaryClass = "astro-ph.CO",
    doi = "10.1103/PhysRevLett.126.171101",
    journal = "Phys. Rev. Lett.",
    volume = "126",
    number = "17",
    pages = "171101",
    year = "2021"
}

@article{Carr:2016drx,
    author = "Carr, Bernard and Kuhnel, Florian and Sandstad, Marit",
    title = "{Primordial Black Holes as Dark Matter}",
    eprint = "1607.06077",
    archivePrefix = "arXiv",
    primaryClass = "astro-ph.CO",
    reportNumber = "NORDITA-2016-83",
    doi = "10.1103/PhysRevD.94.083504",
    journal = "Phys. Rev. D",
    volume = "94",
    number = "8",
    pages = "083504",
    year = "2016"
}

@article{Green:2020jor,
    author = "Green, Anne M. and Kavanagh, Bradley J.",
    title = "{Primordial Black Holes as a dark matter candidate}",
    eprint = "2007.10722",
    archivePrefix = "arXiv",
    primaryClass = "astro-ph.CO",
    doi = "10.1088/1361-6471/abc534",
    journal = "J. Phys. G",
    volume = "48",
    number = "4",
    pages = "043001",
    year = "2021"
}

@article{Fujita:2014hha,
    author = "Fujita, Tomohiro and Kawasaki, Masahiro and Harigaya, Keisuke and Matsuda, Ryo",
    title = "{Baryon asymmetry, dark matter, and density perturbation from primordial black holes}",
    eprint = "1401.1909",
    archivePrefix = "arXiv",
    primaryClass = "astro-ph.CO",
    reportNumber = "IPMU-14-0009, ICRR-REPORT-668-2013-17",
    doi = "10.1103/PhysRevD.89.103501",
    journal = "Phys. Rev. D",
    volume = "89",
    number = "10",
    pages = "103501",
    year = "2014"
}

@article{Allahverdi:2017sks,
    author = "Allahverdi, Rouzbeh and Dent, James and Osinski, Jacek",
    title = "{Nonthermal production of dark matter from primordial black holes}",
    eprint = "1711.10511",
    archivePrefix = "arXiv",
    primaryClass = "astro-ph.CO",
    doi = "10.1103/PhysRevD.97.055013",
    journal = "Phys. Rev. D",
    volume = "97",
    number = "5",
    pages = "055013",
    year = "2018"
}

@article{Lennon:2017tqq,
    author = "Lennon, Olivier and March-Russell, John and Petrossian-Byrne, Rudin and Tillim, Hannah",
    title = "{Black Hole Genesis of Dark Matter}",
    eprint = "1712.07664",
    archivePrefix = "arXiv",
    primaryClass = "hep-ph",
    doi = "10.1088/1475-7516/2018/04/009",
    journal = "JCAP",
    volume = "04",
    pages = "009",
    year = "2018"
}

@article{Morrison:2018xla,
    author = "Morrison, Logan and Profumo, Stefano and Yu, Yan",
    title = "{Melanopogenesis: Dark Matter of (almost) any Mass and Baryonic Matter from the Evaporation of Primordial Black Holes weighing a Ton (or less)}",
    eprint = "1812.10606",
    archivePrefix = "arXiv",
    primaryClass = "astro-ph.CO",
    doi = "10.1088/1475-7516/2019/05/005",
    journal = "JCAP",
    volume = "05",
    pages = "005",
    year = "2019"
}

@article{Hooper:2019gtx,
    author = "Hooper, Dan and Krnjaic, Gordan and McDermott, Samuel D.",
    title = "{Dark Radiation and Superheavy Dark Matter from Black Hole Domination}",
    eprint = "1905.01301",
    archivePrefix = "arXiv",
    primaryClass = "hep-ph",
    reportNumber = "FERMILAB-PUB-19-186-A",
    doi = "10.1007/JHEP08(2019)001",
    journal = "JHEP",
    volume = "08",
    pages = "001",
    year = "2019"
}

@article{Masina:2020xhk,
    author = "Masina, Isabella",
    title = "{Dark matter and dark radiation from evaporating primordial black holes}",
    eprint = "2004.04740",
    archivePrefix = "arXiv",
    primaryClass = "hep-ph",
    doi = "10.1140/epjp/s13360-020-00564-9",
    journal = "Eur. Phys. J. Plus",
    volume = "135",
    number = "7",
    pages = "552",
    year = "2020"
}

@article{Cheek:2021cfe,
    author = "Cheek, Andrew and Heurtier, Lucien and Perez-Gonzalez, Yuber F. and Turner, Jessica",
    title = "{Primordial black hole evaporation and dark matter production. II. Interplay with the freeze-in or freeze-out mechanism}",
    eprint = "2107.00016",
    archivePrefix = "arXiv",
    primaryClass = "hep-ph",
    reportNumber = "FERMILAB-PUB-21-305-T, NUHEP-TH/21-07, CP3-21-42, IPPP/21/01",
    doi = "10.1103/PhysRevD.105.015023",
    journal = "Phys. Rev. D",
    volume = "105",
    number = "1",
    pages = "015023",
    year = "2022"
}

@article{Chattopadhyay:2022fwa,
    author = "Chattopadhyay, Pratik and Chaudhuri, Arnab and Khlopov, Maxim Yu.",
    title = "{Dark matter from evaporating primordial black holes in the early universe}",
    eprint = "2209.11288",
    archivePrefix = "arXiv",
    primaryClass = "hep-ph",
    doi = "10.1142/S0219887824502517",
    journal = "Int. J. Geom. Meth. Mod. Phys.",
    volume = "22",
    number = "01",
    pages = "2450251",
    year = "2025"
}

@article{Baldes:2020nuv,
    author = "Baldes, Iason and Decant, Quentin and Hooper, Deanna C. and Lopez-Honorez, Laura",
    title = "{Non-Cold Dark Matter from Primordial Black Hole Evaporation}",
    eprint = "2004.14773",
    archivePrefix = "arXiv",
    primaryClass = "astro-ph.CO",
    reportNumber = "ULB-TH/20-05",
    doi = "10.1088/1475-7516/2020/08/045",
    journal = "JCAP",
    volume = "08",
    pages = "045",
    year = "2020"
}

@article{Gondolo:2020uqv,
    author = "Gondolo, Paolo and Sandick, Pearl and Shams Es Haghi, Barmak",
    title = "{Effects of primordial black holes on dark matter models}",
    eprint = "2009.02424",
    archivePrefix = "arXiv",
    primaryClass = "hep-ph",
    doi = "10.1103/PhysRevD.102.095018",
    journal = "Phys. Rev. D",
    volume = "102",
    number = "9",
    pages = "095018",
    year = "2020"
}

@article{Boucenna:2017ghj,
    author = "Boucenna, Sofiane M. and Kuhnel, Florian and Ohlsson, Tommy and Visinelli, Luca",
    title = "{Novel Constraints on Mixed Dark-Matter Scenarios of Primordial Black Holes and WIMPs}",
    eprint = "1712.06383",
    archivePrefix = "arXiv",
    primaryClass = "hep-ph",
    doi = "10.1088/1475-7516/2018/07/003",
    journal = "JCAP",
    volume = "07",
    pages = "003",
    year = "2018"
}

@article{Bertone:2019vsk,
    author = "Bertone, Gianfranco and Coogan, Adam M. and Gaggero, Daniele and Kavanagh, Bradley J. and Weniger, Christoph",
    title = "{Primordial Black Holes as Silver Bullets for New Physics at the Weak Scale}",
    eprint = "1905.01238",
    archivePrefix = "arXiv",
    primaryClass = "hep-ph",
    doi = "10.1103/PhysRevD.100.123013",
    journal = "Phys. Rev. D",
    volume = "100",
    number = "12",
    pages = "123013",
    year = "2019"
}

@article{Hertzberg:2020hsz,
    author = "Hertzberg, Mark P. and Schiappacasse, Enrico D. and Yanagida, Tsutomu T.",
    title = "{Axion Star Nucleation in Dark Minihalos around Primordial Black Holes}",
    eprint = "2001.07476",
    archivePrefix = "arXiv",
    primaryClass = "astro-ph.CO",
    doi = "10.1103/PhysRevD.102.023013",
    journal = "Phys. Rev. D",
    volume = "102",
    number = "2",
    pages = "023013",
    year = "2020"
}

@article{Carr:2020mqm,
    author = "Carr, Bernard and Kuhnel, Florian and Visinelli, Luca",
    title = "{Black holes and WIMPs: all or nothing or something else}",
    eprint = "2011.01930",
    archivePrefix = "arXiv",
    primaryClass = "astro-ph.CO",
    doi = "10.1093/mnras/stab1930",
    journal = "Mon. Not. Roy. Astron. Soc.",
    volume = "506",
    number = "3",
    pages = "3648--3661",
    year = "2021"
}

@article{Gines:2022qzy,
    author = "Gin{\'e}s, Estanis Utrilla and Mena, Olga and Witte, Samuel J.",
    title = "{Revisiting constraints on WIMPs around primordial black holes}",
    eprint = "2207.09481",
    archivePrefix = "arXiv",
    primaryClass = "astro-ph.CO",
    doi = "10.1103/PhysRevD.106.063538",
    journal = "Phys. Rev. D",
    volume = "106",
    number = "6",
    pages = "063538",
    year = "2022"
}

@article{Chanda:2022hls,
    author = "Chanda, Prolay and Scholtz, Jakub and Unwin, James",
    title = "{Improved constraints on dark matter annihilations around primordial black holes}",
    eprint = "2209.07541",
    archivePrefix = "arXiv",
    primaryClass = "hep-ph",
    doi = "10.1007/JHEP07(2024)273",
    journal = "JHEP",
    volume = "07",
    pages = "273",
    year = "2024"
}

@article{Bernal:2021bbv,
    author = "Bernal, Nicol{\'a}s and Perez-Gonzalez, Yuber F. and Xu, Yong and Zapata, {\'O}scar",
    title = "{ALP dark matter in a primordial black hole dominated universe}",
    eprint = "2110.04312",
    archivePrefix = "arXiv",
    primaryClass = "hep-ph",
    reportNumber = "PI/UAN-2021-702FT, FERMILAB-PUB-21-478-T, NUHEP-TH/21-16, IPPP/21/36",
    doi = "10.1103/PhysRevD.104.123536",
    journal = "Phys. Rev. D",
    volume = "104",
    number = "12",
    pages = "123536",
    year = "2021"
}

@article{Hawking:1975vcx,
    author = "Hawking, S. W.",
    editor = "Gibbons, G. W. and Hawking, S. W.",
    title = "{Particle Creation by Black Holes}",
    doi = "10.1007/BF02345020",
    journal = "Commun. Math. Phys.",
    volume = "43",
    pages = "199--220",
    year = "1975",
    note = "[Erratum: Commun.Math.Phys. 46, 206 (1976)]"
}

@article{Auffinger:2022khh,
    author = "Auffinger, J{\'e}r{\'e}my",
    title = "{Primordial black hole constraints with Hawking radiation{\textemdash}A review}",
    eprint = "2206.02672",
    archivePrefix = "arXiv",
    primaryClass = "astro-ph.CO",
    doi = "10.1016/j.ppnp.2023.104040",
    journal = "Prog. Part. Nucl. Phys.",
    volume = "131",
    pages = "104040",
    year = "2023"
}

@article{Slatyer:2016qyl,
    author = "Slatyer, Tracy R. and Wu, Chih-Liang",
    title = "{General Constraints on Dark Matter Decay from the Cosmic Microwave Background}",
    eprint = "1610.06933",
    archivePrefix = "arXiv",
    primaryClass = "astro-ph.CO",
    reportNumber = "MIT-CTP-4842",
    doi = "10.1103/PhysRevD.95.023010",
    journal = "Phys. Rev. D",
    volume = "95",
    number = "2",
    pages = "023010",
    year = "2017"
}

@article{Liu:2019bbm,
    author = "Liu, Hongwan and Ridgway, Gregory W. and Slatyer, Tracy R.",
    title = "{Code package for calculating modified cosmic ionization and thermal histories with dark matter and other exotic energy injections}",
    eprint = "1904.09296",
    archivePrefix = "arXiv",
    primaryClass = "astro-ph.CO",
    doi = "10.1103/PhysRevD.101.023530",
    journal = "Phys. Rev. D",
    volume = "101",
    number = "2",
    pages = "023530",
    year = "2020"
}

@article{Fields:2019pfx,
    author = "Fields, Brian D. and Olive, Keith A. and Yeh, Tsung-Han and Young, Charles",
    title = "{Big-Bang Nucleosynthesis after Planck}",
    eprint = "1912.01132",
    archivePrefix = "arXiv",
    primaryClass = "astro-ph.CO",
    reportNumber = "UMN--TH--3902/19, FTPI--MINN--19/25",
    doi = "10.1088/1475-7516/2020/03/010",
    journal = "JCAP",
    volume = "03",
    pages = "010",
    year = "2020",
    note = "[Erratum: JCAP 11, E02 (2020)]"
}

@article{Keith:2022sow,
    author = "Keith, Celeste and Hooper, Dan and Linden, Tim and Liu, Rayne",
    title = "{Sensitivity of future gamma-ray telescopes to primordial black holes}",
    eprint = "2204.05337",
    archivePrefix = "arXiv",
    primaryClass = "astro-ph.HE",
    reportNumber = "FERMILAB-PUB-22-251-T",
    doi = "10.1103/PhysRevD.106.043003",
    journal = "Phys. Rev. D",
    volume = "106",
    number = "4",
    pages = "043003",
    year = "2022"
}

@article{Tashiro:2008sf,
    author = "Tashiro, Hiroyuki and Sugiyama, Naoshi",
    title = "{Constraints on Primordial Black Holes by Distortions of Cosmic Microwave Background}",
    eprint = "0801.3172",
    archivePrefix = "arXiv",
    primaryClass = "astro-ph",
    doi = "10.1103/PhysRevD.78.023004",
    journal = "Phys. Rev. D",
    volume = "78",
    pages = "023004",
    year = "2008"
}

@article{Acharya:2019xla,
    author = "Acharya, Sandeep Kumar and Khatri, Rishi",
    title = "{CMB spectral distortions constraints on primordial black holes, cosmic strings and long lived unstable particles revisited}",
    eprint = "1912.10995",
    archivePrefix = "arXiv",
    primaryClass = "astro-ph.CO",
    doi = "10.1088/1475-7516/2020/02/010",
    journal = "JCAP",
    volume = "02",
    pages = "010",
    year = "2020"
}

@article{Escudero:2023vgv,
    author = "Escudero, Miguel and Pooni, Charis Kaur and Fairbairn, Malcolm and Blas, Diego and Du, Xiaolong and Marsh, David J. E.",
    title = "{Axion star explosions: A new source for axion indirect detection}",
    eprint = "2302.10206",
    archivePrefix = "arXiv",
    primaryClass = "hep-ph",
    reportNumber = "KCL-PH-TH-2023-16, CERN-TH-2023-029",
    doi = "10.1103/PhysRevD.109.043018",
    journal = "Phys. Rev. D",
    volume = "109",
    number = "4",
    pages = "043018",
    year = "2024"
}

@article{Kodama:1982sf,
    author = "Kodama, Hideo and Sasaki, Misao and Sato, Katsuhiko",
    title = "{Abundance of Primordial Holes Produced by Cosmological First Order Phase Transition}",
    reportNumber = "KUNS 642",
    doi = "10.1143/PTP.68.1979",
    journal = "Prog. Theor. Phys.",
    volume = "68",
    pages = "1979",
    year = "1982"
}

@article{Hawking:1982ga,
    author = "Hawking, S. W. and Moss, I. G. and Stewart, J. M.",
    title = "{Bubble Collisions in the Very Early Universe}",
    reportNumber = "Print-82-0180 (CAMBRIDGE)",
    doi = "10.1103/PhysRevD.26.2681",
    journal = "Phys. Rev. D",
    volume = "26",
    pages = "2681",
    year = "1982"
}

@article{Bertone:2024wbn,
    author = "Bertone, Gianfranco and Wierda, A. Renske A. C. and Gaggero, Daniele and Kavanagh, Bradley J. and Volonteri, Marta and Yoshida, Naoki",
    title = "{Toward a realistic description of dark matter overdensities around black holes}",
    eprint = "2404.08731",
    archivePrefix = "arXiv",
    primaryClass = "astro-ph.CO",
    doi = "10.1103/5nnf-8fz9",
    journal = "Phys. Rev. D",
    volume = "112",
    number = "4",
    pages = "043537",
    year = "2025"
}

@article{Jangra:2024sif,
    author = "Jangra, Pratibha and Gaggero, Daniele and Kavanagh, Bradley J. and Diego, J. M.",
    title = "{The cosmic history of Primordial Black Hole accretion and its uncertainties}",
    eprint = "2412.11921",
    archivePrefix = "arXiv",
    primaryClass = "astro-ph.CO",
    doi = "10.1088/1475-7516/2025/08/006",
    journal = "JCAP",
    volume = "08",
    pages = "006",
    year = "2025"
}

@article{Page:1976wx,
    author = "Page, Don N. and Hawking, S. W.",
    title = "{Gamma rays from primordial black holes}",
    doi = "10.1086/154350",
    journal = "Astrophys. J.",
    volume = "206",
    pages = "1--7",
    year = "1976"
}

@article{Arbey:2021mbl,
    author = "Arbey, Alexandre and Auffinger, J{\'e}r{\'e}my",
    title = "{Physics Beyond the Standard Model with BlackHawk v2.0}",
    eprint = "2108.02737",
    archivePrefix = "arXiv",
    primaryClass = "gr-qc",
    reportNumber = "CERN-TH-2021-117",
    doi = "10.1140/epjc/s10052-021-09702-8",
    journal = "Eur. Phys. J. C",
    volume = "81",
    pages = "910",
    year = "2021"
}

@article{Carr:1976zz,
    author = "Carr, Bernard J.",
    title = "{Some cosmological consequences of primordial black-hole evaporations}",
    doi = "10.1086/154351",
    journal = "Astrophys. J.",
    volume = "206",
    pages = "8--25",
    year = "1976"
}

@article{MacGibbon:1990zk,
    author = "MacGibbon, J. H. and Webber, B. R.",
    title = "{Quark and gluon jet emission from primordial black holes: The instantaneous spectra}",
    doi = "10.1103/PhysRevD.41.3052",
    journal = "Phys. Rev. D",
    volume = "41",
    pages = "3052--3079",
    year = "1990"
}

@article{MacGibbon:1991tj,
    author = "MacGibbon, Jane H.",
    title = "{Quark and gluon jet emission from primordial black holes. 2. The Lifetime emission}",
    reportNumber = "LHEA-91-001",
    doi = "10.1103/PhysRevD.44.376",
    journal = "Phys. Rev. D",
    volume = "44",
    pages = "376--392",
    year = "1991"
}

@article{Teukolsky:1973ha,
    author = "Teukolsky, Saul A.",
    title = "{Perturbations of a rotating black hole. 1. Fundamental equations for gravitational electromagnetic and neutrino field perturbations}",
    doi = "10.1086/152444",
    journal = "Astrophys. J.",
    volume = "185",
    pages = "635--647",
    year = "1973"
}

@article{Bardeen:1973xb,
    author = "Bardeen, J. M. and Press, W. H.",
    title = "{Radiation fields in the schwarzschild background}",
    doi = "10.1063/1.1666175",
    journal = "J. Math. Phys.",
    volume = "14",
    pages = "7--19",
    year = "1973"
}

@article{Boccia:2024nly,
    author = "Boccia, Andrea and Iocco, Fabio and Visinelli, Luca",
    title = "{Constraining the primordial black hole abundance through big-bang nucleosynthesis}",
    eprint = "2405.18493",
    archivePrefix = "arXiv",
    primaryClass = "astro-ph.CO",
    doi = "10.1103/PhysRevD.111.063508",
    journal = "Phys. Rev. D",
    volume = "111",
    number = "6",
    pages = "063508",
    year = "2025"
}

\end{document}